\documentclass[journal]{IEEEtran}
\usepackage{cite}

\ifCLASSINFOpdf
\else
\fi
\usepackage{graphicx}
\usepackage{caption}
\usepackage{subcaption}
\usepackage{multirow}
\usepackage{amsmath, amssymb, etoolbox}
\usepackage{algorithmic}

\usepackage{array}
\usepackage{makecell}
\usepackage{xcolor}	
\usepackage{color, colortbl}
\usepackage{comment}
\usepackage{adjustbox}
 \usepackage[usestackEOL]{stackengine}
\DeclareCaptionFormat{myformat}{\fontsize{8}{8}\selectfont#1#2#3}
\usepackage{nomencl}
\makenomenclature
\renewcommand\nomgroup[1]{%
  \item{\bfseries \ifstrequal{#1}{S}{List of Sets}{%
  \ifstrequal{#1}{V}{List of Variables}{%
  \ifstrequal{#1}{O}{Other symbols}{}}}}}

\begin{document}
\bstctlcite{bstctl:nodash}
%
\title{Optimal Placement of Smart Hybrid Transformers in Distribution Networks\thanks{This work has been submitted to the IEEE for possible publication. Copyright may be transferred without notice, after which this version may no longer be accessible.}}
%
%
%

\author{Samuel Hayward,\thanks{Samuel Hayward is with the University of Edinburgh, Edinburgh, United Kingdom (email: s1547241@ed.ac.uk).} \and
Martin Doff-Sotta,\thanks{Martin Doff-Sotta is with the University of Oxford, Oxford, United Kingdom (email: martin.doff-sotta@eng.ox.ac.uk).}
        Michael Merlin,~\IEEEmembership{Member, IEEE,}\thanks{Michael Merlin is with the University of Edinburgh, Edinburgh, United Kingdom (email: michael.merlin@ed.ac.uk).} \and
        Matthew Williams,\thanks{Mattew Williams is with the private company IONATE, London, United Kingdom (email: matthew@ionate.energy).} \and
        and Thomas Morstyn,~\IEEEmembership{Senior Member, IEEE}\thanks{Thomas Morstyn is with the University of Oxford, Oxford, United Kingdom (email: thomas.morstyn@eng.ox.ac.uk).}
        }
\maketitle
\let\thefootnote\relax\footnotetext{This work was supported by the Scottish Government and IONATE through the ETP Energy Industry Doctorate Programme (project reference 180) and by the Engineering and Physical Sciences Research Council (EPSRC) (project reference EP/X027384/1).}
\begin{abstract}
Hybrid transformers are a relatively new technology that combine conventional power transformers with power electronics to provide voltage and reactive power control capabilities in distribution networks. This paper proposes a novel method of determining the optimal location and utilisation of hybrid transformers in 3-phase distribution networks to maximise the net present value of hybrid transformers based on their ability to increase the export of power produced by distributed generators over their operational lifespan. This has been accomplished through sequential linear programming, a key feature of which is the consideration of nonlinear characteristics and constraints relating to hybrid transformer power electronics and control capabilities. Test cases were carried out in a modified version of the Cigre European Low Voltage Distribution Network Benchmark, which has been extended by connecting it with two additional low voltage distribution test networks. All test case results demonstrate that the installation and utilisation of hybrid transformers can improve the income earned from exporting excess active power, justifying their installation cost (with the highest net present value being £6.56 million, resulting from a 45.53\% increase in estimated annual profits due to coordinated HT compensation).
\end{abstract}

\begin{IEEEkeywords}
hybrid transformer, optimal placement, optimal power flow, power injection model, power quality, renewable distributed generation, sequential linear programming, successive linear programming, voltage control.
\end{IEEEkeywords}

%
\IEEEpeerreviewmaketitle

\section{INTRODUCTION.}
\label{section:Intro}
%
%
%
%
\IEEEPARstart{O}{ver} the past few years, there has been a greater effort from governments around the world to combat climate change by reducing green house gas emissions. This has motivated the rollout of renewable distributed generators (DG) in electrical distribution networks \cite{Bala2012,Burkard2015,Zheng2022,Foti2020}. As an example, the global capacity of distributed solar photovoltaic (PV) systems has increased by 476 GW between the years 2018 and 2023, and is predicted to increase by an additional 1,659 GW by 2030 \cite{IEA_RENEWABLES2024}. However, using renewable DGs can lead to power quality issues which can impact the reliability of the network, such as poor power factors and voltage fluctuations \cite{Bala2012,Burkard2015,Foti2020}. It is possible to alleviate such power quality issues by reinforcing distribution networks, however, this is expensive and time-consuming \cite{Bala2012}. Alternatively, volt/VAR control devices can be added to distribution networks \cite{Burkard2020}. Devices that can help mitigate the power quality issues caused by renewable DG include transformers with On-Load Tap Changers (OLTC), Solid-State Transformers (SST), and Hybrid Transformers (HT) \cite{Burkard2020,Winter2019}.

An OLTC is able to control a transformer output voltage by making adjustments to its turns ratio \cite{Eremia2013}. However, frequent tap position switching causes wear and tear \cite{Divan2016,Katiraei2011}, and it can only swap between a limited number of discrete voltage levels. These characteristics make it difficult for OLTCs to effectively manage renewable DG fluctuations \cite{Burkard2020}.

An SST is a power electronic device that is capable of 100\% voltage regulation and full active and reactive power flow control \cite{ Zheng2022,Huber2019}. However, compared to conventional power transformers, SSTs typically have shorter operational lifespans, are sensitive to sudden and large increases in voltage and current,  and are expensive due to the high power rating of their power electronic converters\cite{Bala2012,Zheng2022,Huber2019,Liserre2016, Mollik2022}.

An HT is a conventional power transformer integrated with fractionally rated power electronics \cite{Bala2012,Burkard2015}. The addition of power electronics provides the HT with control capabilities such as dynamic voltage regulation and power factor correction \cite{Bala2012}. HTs offer less controllability than SSTs because their power electronics are only rated for a fraction of the transformer's total transferable power (generally 10 to 20\%), whereas SST power electronics are rated for the full transferable power \cite{Bala2012,Carreno2021}. Despite this, the control capabilities of an HT can still be sufficient to mitigate power quality issues. Furthermore, if the power electronics fail, an HT can still operate as a conventional transformer, whereas SSTs cannot \cite{Bala2012,Burkard2015}. Compared to conventional transformers (which have an efficiency of 99\%), HTs are less efficient, with an efficiency range of 98.6-98.8\%, due to the addition of power electronics \cite{Zheng2022,Huber2019}. However, HTs are more efficient than SSTs which generally have lower efficiencies of around 97\% \cite{Zheng2022,Huber2019,Liserre2016}. Due to the inclusion of power electronics, both HTs and SSTs have higher material costs than conventional power transformers \cite{Zheng2022}. However, since HT power electronics are fractionally rated as opposed to fully rated, HTs are less expensive than SSTs (which can potentially be 5 to 25 times more expensive than conventional power transformers) \cite{Bala2012,Zheng2022,Huber2019,Mollik2022}. A disadvantage of HTs is their volume. HTs, compared to conventional power transformers and SSTs of similar power ratings, tend to have a larger volume \cite{Zheng2022}. Despite this, compared to OLTCs and SSTs, HTs remain an overall appealing choice for handling power quality issues caused by renewable DGs, offering a balance of efficiency, cost, robustness, and controllability.

Various methods exist for optimally coordinating and/or placing volt/VAR devices to control network voltage levels and facilitate greater DG installation. For example, \cite{BazrafshanOPF2019} describes an optimal power flow (OPF) method which is able to coordinate the tap positions of step-voltage regulators (SVR), \cite{Zhao2016} describes an OPF method for active distribution networks that coordinates OLTC tap positions and reactive power compensators, and \cite{Nusair2017} presents an algorithm for minimising network power losses by coordinating OLTC tap positions, capacitor banks, and Static Synchronous Compensators (STATCOM), while also determining the optimal location for STATCOM devices. Yet, there are very few OPF methods that accommodate HTs, such as \cite{Gao2021}, which introduces a bi-level programming approach for HT allocation and control to minimise network power losses.  However, in this study, only the reactive power compensation capability of HTs is considered and the voltage regulation capability of HTs is not utilised. In \cite{Hayward2024}, the authors presented a Sequential Linear Programming (SLP) method which determines the optimal use of HTs in a distribution network to maximise DG output, utilising both the voltage regulation and reactive power compensation capabilities of HTs. However, the SLP method in \cite{Hayward2024} assumes a fixed location for the HTs and does not optimise their position. Therefore, there is a lack of optimisation methods that can both determine the optimal locations of multiple HTs in distribution networks and make full use of HT control capabilities. Developing such methods would be valuable for network operators considering the installation of HTs to enable greater DG export capacity.

The main contribution of this paper is the development of an SLP method that determines the optimal locations for HTs within a distribution network in order to maximise the Net Present Value (NPV) of HTs based on the additional income earned by exporting excess power generated by DGs over the operational life span of the HTs. The proposed SLP method is referred to as the HT SLP in this paper. The HT SLP is capable of coordinating multiple HT models within the same distribution network. The proposed HT SLP method is capable of addressing the nonlinear constraints and characteristics of distribution networks and HTs by solving a series of Linear Programming (LP) problems iteratively, avoiding the need to solve a Nonlinear Program (NLP). Furthermore, the HT SLP method is able to work with unbalanced, multi-phase distribution network models. The proposed HT SLP was tested in a case study with a combination of the Cigre Low Voltage (LV) distribution network benchmark and two other LV network benchmarks. Test case results show that the proposed HT SLP is capable of utilising HTs to improve the income earned from exporting excess active power, resulting in positive NPVs. The highest NPV earned was £6.56 million, which resulted from a 45.53\% increase in estimated annual profits due to coordinated HT compensation.

In this paper, Section \ref{section:HTPIM} presents steady-state models of HTs. Section \ref{section:SLP} describes the HT SLP method. Section \ref{section:Results} describes the test case network and presents the results, with positive NPVs demonstrating that the HT SLP is capable of optimally placing and coordinating HTs. Section \ref{section:Conclusion} concludes the paper.

For this paper, $\mathcal{N}$ is the set of network buses (excluding the grid-connected slack bus), $\mathcal{N}^{\mathrm{Y}}$ is the set of wye-connected network buses (subset of $\mathcal{N}$), $\mathcal{N}^{\Delta}$ is the set of delta-connected network buses (subset of $\mathcal{N}$), $\mathcal{N}^\mathrm{{DG}}$ is the set of network buses that have DG units, $\mathcal{N}^\mathrm{{HTp}}$ is the set of HT virtual buses, $\mathcal{N}^\mathrm{{HTs}}$ is the set of HT LV buses, $\mathcal{N}^\mathrm{t}$ is the set of time periods, $\mathcal{\varepsilon}^\mathrm{{HT}}$ is the set of lines between HT virtual buses and LV buses (where $(i,j)$ denotes a line from virtual bus $i$ to LV bus $j$), and $\Omega$ is the set of network phases (e.g., 0, 1, 2).

\section{HYBRID TRANSFORMER MODEL.}
\label{section:HTPIM}
The proposed HT SLP uses the volt/VAR control capabilities of HTs to prevent voltage violations that can be caused by increasing DG active power exports. This can be accomplished through static network modelling. As such, this section presents quasi-static models used to represent HTs in the HT SLP.

\begin{figure}[t!]
    \centering
    \includegraphics[width=0.8\linewidth]{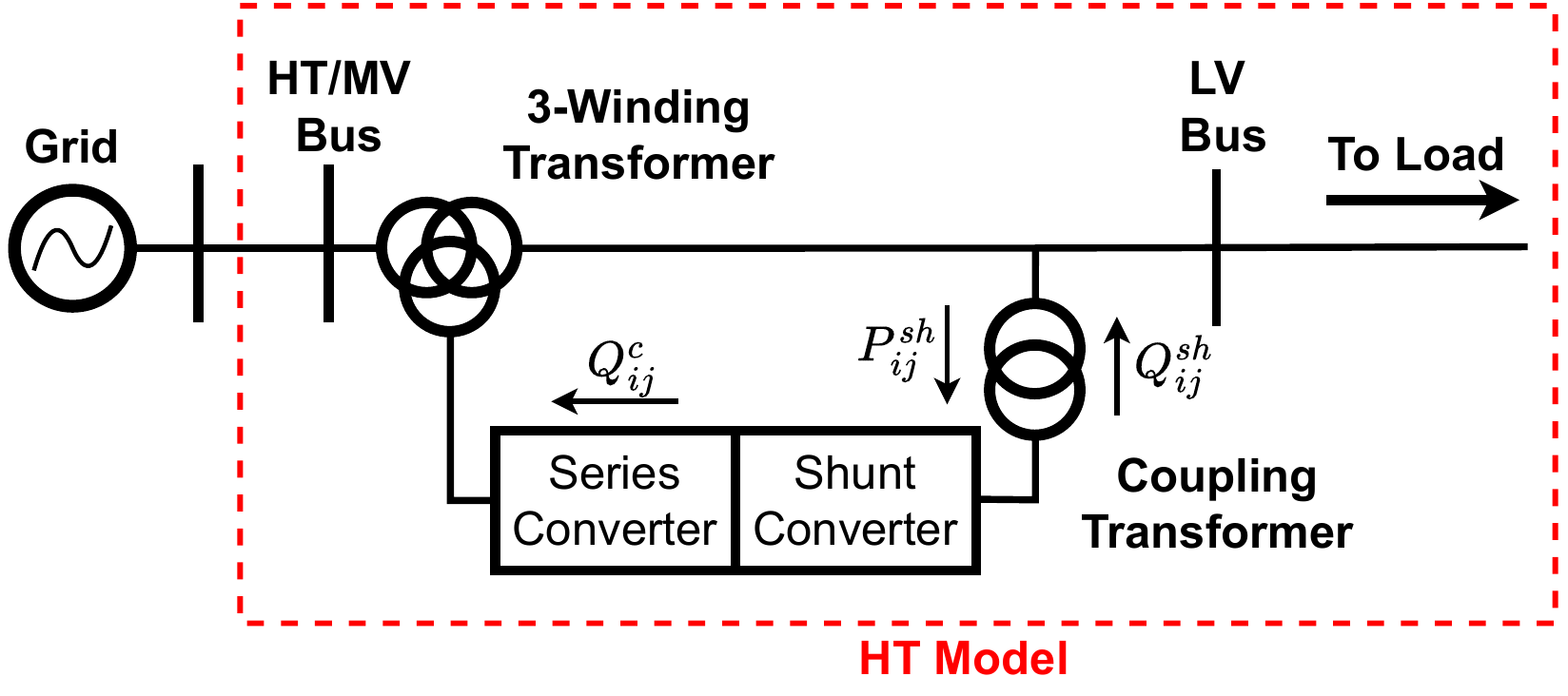}
    \caption{Single line diagram of the three-winding transformer HT model.}
    \label{fig:HT_tf}
\end{figure}

There exist multiple different topologies for HTs \cite{Burkard2015, Carreno2021}. In this paper, it has been assumed that the HT is a three-winding power transformer connected to a pair of power electronic converters, where one converter is connected to the transformer in series through the tertiary winding (referred to as the series converter) and the other converter connects to the output of the transformer through a shunt connection (referred to as the shunt converter). A diagram of this HT topology is shown in Figure \ref{fig:HT_tf}. This HT topology has been selected because it is able to provide both voltage compensation and reactive power compensation. This HT can only generate reactive power. Active power cannot be generated, but can be transferred through the power electronic converters. The amount of reactive power that can be generated is limited by both the maximum apparent power rating of the power electronic converters and the active power being exchanged between the converters.

\begin{figure}[t!]
    \centering
    \includegraphics[width=0.8\linewidth]{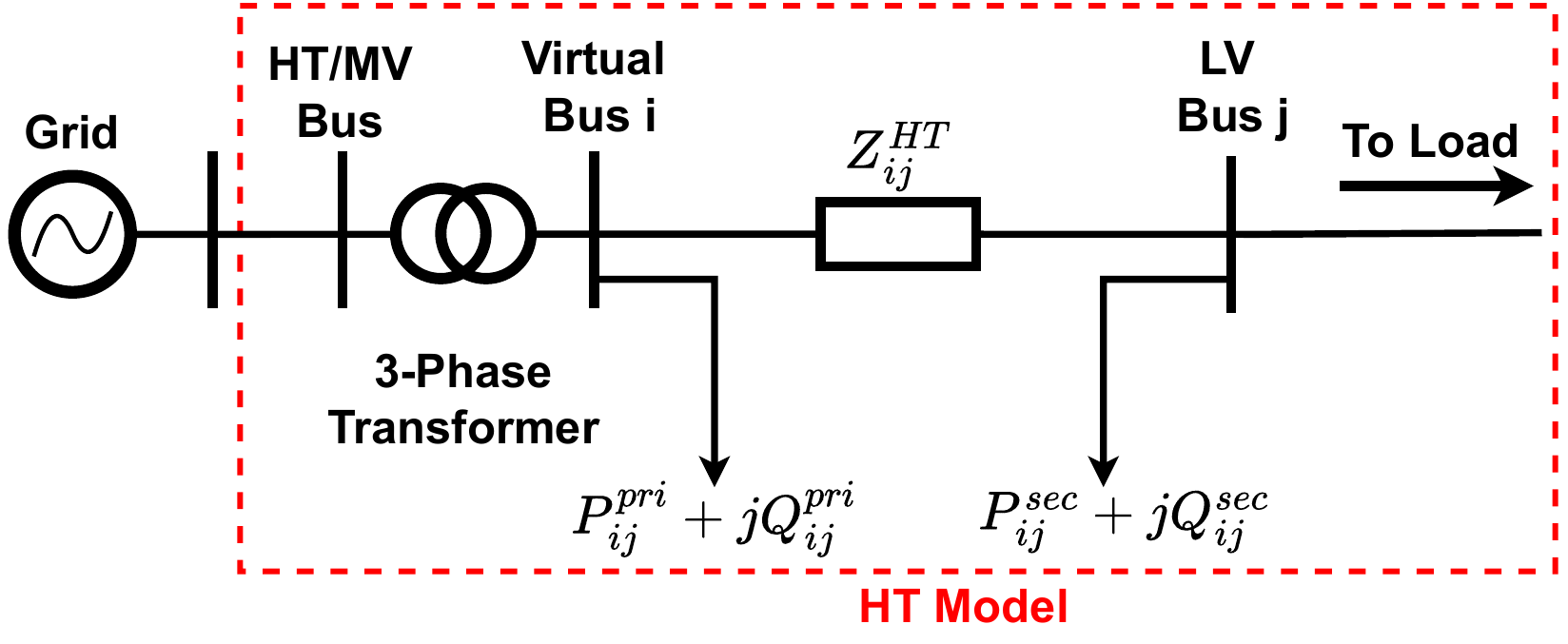}
    \caption{Single line diagram of the HT PIM.}
    \label{fig:HT_pim}
\end{figure}

The HT can be represented with the Power Injection Model (PIM) in Figure \ref{fig:HT_pim}, where the volt/VAR control capabilities of the HT are modelled by controlled power injections. The HT PIM equations are given as follows \cite{Hayward2024}
\begingroup
\allowdisplaybreaks
\begin{align}\underline{r} &\leq r_{ij,\phi} \leq \overline{r} \label{eq:nonlin_r},\\
-\pi &\leq \gamma_{ij,\phi} \leq \pi \label{eq:nonlin_gamma},\\
e_{ij,\phi}^\mathrm{{pq}} &= Re(r_{ij,\phi}v_{i,\phi}exp(-j\gamma_{ij,\phi})), \label{eq: nonlin_epq}\\
f_{ij,\phi}^\mathrm{{pq}} &= Im(r_{ij,\phi}v_{i,\phi}exp(-j\gamma_{ij,\phi})),\label{eq: nonlin_fpq}\\
P_{ij,\phi}^\mathrm{{pri}} &= \begin{aligned}[t]&G_{ij,\phi\phi}(-e_{i, \phi}e_{ij,\phi}^\mathrm{{pq}}-f_{i, \phi}f_{ij,\phi}^\mathrm{{pq}})\\&+B_{ij,\phi\phi}(e_{i, \phi}f_{ij,\phi}^\mathrm{{pq}}-f_{i, \phi}e_{ij,\phi}^\mathrm{{pq}}),
\end{aligned} \label{eq: nonlin_Ppri}\\
Q_{ij,\phi}^\mathrm{{pri}} &= \begin{aligned}[t]&G_{ij,\phi\phi}(e_{i, \phi}f_{ij,\phi}^\mathrm{{pq}}-f_{i, \phi}e_{ij,\phi}^\mathrm{{pq}})\\&+B_{ij,\phi\phi}(e_{i, \phi}e_{ij,\phi}^\mathrm{{pq}}+f_{i, \phi}f_{ij,\phi}^\mathrm{{pq}}),
\end{aligned}\label{eq: nonlin_Qpri}\\
P_{ij,\phi}^\mathrm{{se}} &= \begin{aligned}[t]&G_{ij,\phi\phi}(e_{j, \phi}e_{ij,\phi}^\mathrm{{pq}}+f_{j, \phi}f_{ij,\phi}^\mathrm{{pq}})\\&+B_{ij,\phi\phi}(f_{j, \phi}e_{ij,\phi}^\mathrm{{pq}}-e_{j, \phi}f_{ij,\phi}^\mathrm{pq}),
\end{aligned}\label{eq: nonlin_Pse}\\
Q_{ij,\phi}^\mathrm{se} &= \begin{aligned}[t]&G_{ij,\phi\phi}(f_{j, \phi}e_{ij,\phi}^\mathrm{pq}-e_{j, \phi}f_{ij,\phi}^\mathrm{pq})\\&-B_{ij,\phi\phi}(e_{j, \phi}e_{ij,\phi}^\mathrm{pq}+f_{j, \phi}f_{ij,\phi}^\mathrm{pq}),
\end{aligned}\label{eq: nonlin_Qse}\\
P_{ij,\phi}^\mathrm{sh} &= \begin{aligned}[t]&G_{ij,\phi\phi}\biggl(\left(e_{ij,\phi}^\mathrm{pq}\right)^2+\left( f_{ij,\phi}^\mathrm{pq} \right)^2\\ &+e_{i, \phi}e_{ij,\phi}^\mathrm{pq}-e_{j, \phi}e_{ij,\phi}^\mathrm{pq} +f_{i, \phi}f_{ij,\phi}^\mathrm{pq}\\ &- f_{j, \phi}f_{ij,\phi}^\mathrm{pq}\biggr) +B_{ij,\phi\phi}(e_{i, \phi}f_{ij,\phi}^\mathrm{pq}\\ &-e_{j, \phi}f_{ij,\phi}^\mathrm{pq}-f_{i, \phi}e_{ij,\phi}^\mathrm{pq}+f_{j, \phi}e_{ij,\phi}^\mathrm{pq}),
\end{aligned}\label{eq: nonlin_Psh}\\
Q_{ij,\phi}^\mathrm{c} &=\begin{aligned}[t]&G_{ij,\phi\phi}(e_{i, \phi}f_{ij,\phi}^\mathrm{pq}-e_{j, \phi}f_{ij,\phi}^\mathrm{pq}-f_{i, \phi}e_{ij,\phi}^\mathrm{pq}\\&+f_{j, \phi}e_{ij,\phi}^\mathrm{pq}) -B_{ij,\phi\phi}\biggl(\left(e_{ij,\phi}^\mathrm{pq}\right)^2\\&+\left( f_{ij,\phi}^\mathrm{pq} \right)^2+e_{i, \phi}e_{ij,\phi}^\mathrm{pq}-e_{j, \phi}e_{ij,\phi}^\mathrm{pq}\\ &+f_{i, \phi}f_{ij,\phi}^\mathrm{pq}- f_{j, \phi}f_{ij,\phi}^\mathrm{pq}\biggr),
\end{aligned}\label{eq: nonlin_Qconv}\\
\lvert \overline{Q}_{ij,\phi}\rvert &= \begin{aligned}[t]
\left(\left(\overline{S}_{ij,\phi}\right)^2 -\left(P_{ij,\phi}^\mathrm{sh}\right)^2\right)^{\frac{1}{2}},
\end{aligned}\label{eq: nonlin_Qmax}\\
-\lvert \overline{Q}_{ij,\phi}\rvert &\leq \begin{aligned}[t]&Q_{ij,\phi}^\mathrm{sh} \leq \lvert \overline{Q}_{ij,\phi}\rvert,
\end{aligned}\label{eq: nonlin_Qsh}\\
-\lvert \overline{Q}_{ij,\phi}\rvert &\leq \begin{aligned}[t]&Q_{ij,\phi}^\mathrm{c} \leq \lvert \overline{Q}_{ij,\phi}\rvert,
\end{aligned}\label{eq: nonlin_Qconv_lim}\\
P_{ij,\phi}^\mathrm{sec} &= P_{ij,\phi}^\mathrm{se}-P_{ij,\phi}^\mathrm{sh},\label{eq: nonlin_Psec}\\
Q_{ij,\phi}^\mathrm{sec} &= Q_{ij,\phi}^\mathrm{se}+Q_{ij,\phi}^\mathrm{sh}.\label{eq: nonlin_Qsec}
\end{align}
\endgroup
where $v \in \mathbb{C}^{3\mathcal{N}}$ are the complex phase voltages of the network ($v_{i,\phi}$ representing the voltage for phase $\phi$ of bus $i$). $e,f \in \mathbb{R}^{3\mathcal{N}}$ represent the real and imaginary parts of $v$. $e_{ij}^\mathrm{pq},f_{ij}^\mathrm{pq} \in \mathbb{R}^{\Omega}$ respectively represent the real and imaginary compensation voltage provided by the HT with virtual bus $i$ and LV bus $j$. $r_{ij} \in \mathbb{R}^{\Omega}$ is the coefficient that defines the magnitude of HT compensating voltage relative to $v_i$, and $\underline{r},\overline{r} \in \mathbb{R}$ are the minimum and maximum coefficient values. $\gamma_{ij} \in \mathbb{R}^{\Omega}$ contains the phase shift angles of the HT compensating voltages. $B_{ij}, G_{ij} \in \mathbb{R}^{\Omega\times \Omega}$ are the susceptance and conductance matrices for the HT line between virtual bus $i$ and LV bus $j$. $P^\mathrm{pri}_{ij}, Q^\mathrm{pri}_{ij} \in \mathbb{R}^{\Omega}$ are the active and reactive power injections made at HT virtual bus $i$. $P^\mathrm{se}_{ij}, Q^\mathrm{se}_{ij} \in \mathbb{R}^{\Omega}$ are the active and reactive power injections made at HT LV bus $j$ due to voltage compensation. $P^\mathrm{sh}_{ij} \in \mathbb{R}^{\Omega}$ is the active power exchange between the shunt and series converters, represented as an active power injection at LV bus $j$. $Q^\mathrm{c}_{ij} \in \mathbb{R}^{\Omega}$ is the reactive power in the series converter. $\lvert\overline{Q}_{ij}\rvert \in \mathbb{R}^{\Omega}$ is the magnitude of the maximum reactive power that can be present in the power electronic converters. $\overline{S}_{ij} \in \mathbb{R}$ is the maximum apparent power rating for the HT power electronic converters, and $Q_{ij}^\mathrm{sh} \in \mathbb{R}^{\Omega}$ is the reactive power provided by the shunt converter at HT LV bus $j$. $P^\mathrm{sec}_{ij}, Q^\mathrm{sec}_{ij} \in \mathbb{R}^{\Omega}$ are the overall active and reactive power injections made at HT LV bus $j$.  It should be noted that (\ref{eq: nonlin_Ppri})-(\ref{eq: nonlin_Qconv}) can be combined and rewritten to provide $P_{ij}^\mathrm{pri}$, $Q_{ij}^\mathrm{pri}$, $P_{ij}^\mathrm{se}$, $Q_{ij}^\mathrm{se}$, $P_{ij}^\mathrm{sh}$, and $Q_{ij}^\mathrm{c}$ in the form of apparent power as $S_{ij}^\mathrm{pri}$, $S_{ij}^\mathrm{se} \in \mathbb{C}^{\Omega}$, and $S_{ij}^\mathrm{c} \in \mathbb{R}^{\Omega}$.

\begingroup
\allowdisplaybreaks
\begin{align}
    v_{ij,\phi}^\mathrm{pq} &= e_{ij,\phi}^\mathrm{pq}+jf_{ij,\phi}^\mathrm{pq}, \label{eq: nonlin_vpq}\\
    S_{ij, \phi}^\mathrm{pri} &= P_{ij,\phi}^\mathrm{pri} + jQ_{ij,\phi}^\mathrm{pri} = \begin{aligned}
        v_{i,\phi}\left(\frac{-v_{ij,\phi}^\mathrm{pq}}{Z_{ij,\phi\phi}^\mathrm{HT}}\right)^{*},\label{eq: nonlin_Spri}
    \end{aligned}\\
    S_{ij, \phi}^\mathrm{se} &= P_{ij,\phi}^\mathrm{se} + jQ_{ij,\phi}^\mathrm{se} = \begin{aligned}
        v_{j,\phi}\left(\frac{v_{ij,\phi}^\mathrm{pq}}{Z_{ij,\phi\phi}^\mathrm{HT}}\right)^{*}\label{eq: nonlin_Sse},
    \end{aligned}\\
    S_{ij, \phi}^\mathrm{c} &= \sqrt{\left(P_{ij,\phi}^\mathrm{sh}\right)^{2} + \left(Q_{ij,\phi}^\mathrm{c}\right)^{2}}, \label{eq: nonlin_Sconv} \\ \nonumber &= \begin{aligned}\left\lvert v_{ij,\phi}^\mathrm{pq}\left(\frac{v_{i,\phi}+v_{ij,\phi}^\mathrm{pq}-v_{j,\phi}}{Z_{ij, \phi\phi}^\mathrm{HT}}\right)^* \right\rvert.
    \end{aligned}
\end{align}
\endgroup
where $v^\mathrm{pq}_{ij} \in \mathbb{C}^{\Omega}$ is the complex compensating voltage provided by the HT. $Z_{ij}^\mathrm{HT} \in \mathbb{C}^{\Omega \times \Omega}$ is the impedance matrix for the line between virtual bus $i$ and LV bus $j$, representing the transformer tertiary winding equivalent impedance.

The three-phase transformer does not actively contribute to the volt/VAR control of the HT, instead, it is a passive component. Nodal admittance matrices, defined by the three-phase transformer's leakage admittance, are used to model the transformer \cite{Das2016}. A delta/wye-g winding configuration is used for the three-phase transformers for all of the HTs presented in this paper. Equations (\ref{eq: tf_mat1}) and (\ref{eq: tf_mat2}) define the nodal admittance matrix for a delta/wye-g three-phase transformer, where $\overline{Y}_\mathrm{Node} \in \mathbb{C}^{6 \times 6}$ is the nodal admittance matrix, $Y_{I}, Y_{II}, Y_{II} \in \mathbb{C}^{3 \times 3}$ are submatrices of $\overline{Y}_\mathrm{Node}$, and $y_\mathrm{t} \in \mathbb{C}$ is the per phase transformer leakage admittance. By utilising the nodal admittance matrix, the three-phase transformer will be treated as a transmission line. This method accounts for power losses in the three-phase transformer.

\begin{align}
\overline{Y}_\mathrm{Node} =
\begin{bmatrix}
\overline{Y}_{II}&\overline{Y}_{III}\\
\overline{Y}_{III}&\overline{Y}_{I}
\end{bmatrix} \label{eq: tf_mat1}
\end{align}
\begin{align}
\allowdisplaybreaks
\overline{Y}_{I} = 
\begin{bmatrix}
y_\mathrm{t}&0&0\\
0&y_\mathrm{t}&0\\
0&0&y_\mathrm{t}
\end{bmatrix}, \overline{Y}_{II} = \frac{1}{3}
\begin{bmatrix}
2y_\mathrm{t}&-y_\mathrm{t}&-y_\mathrm{t}\\
-y_\mathrm{t}&2y_\mathrm{t}&-y_\mathrm{t}\\
-y_\mathrm{t}&-y_\mathrm{t}&2y_\mathrm{t}
\end{bmatrix}, \label{eq: tf_mat2} \\ \nonumber \overline{Y}_{III} = \frac{1}{\sqrt{3}}
\begin{bmatrix}
-y_\mathrm{t}&y_\mathrm{t}&0\\
0&-y_\mathrm{t}&y_\mathrm{t}\\
y_\mathrm{t}&0&-y_\mathrm{t}
\end{bmatrix}.
\end{align} 

\section{SEQUENTIAL LINEAR PROGRAMMING METHOD.}
\label{section:SLP}
This section describes the proprosed HT SLP method. SLP is a method of solving NLP problems, in which the NLP is approximated with a sequence of LP problems which are solved iteratively. Inaccuracies due to linear approximations are minimised by applying restrictions to LP constraints, which limit how far the LPs can deviate from their initialisation points. These restrictions are updated in response to LP results (e.g., inaccurate results will cause restrictions to be tightened). LP results are also used to update initialisation points for the next iteration of the SLP. The SLP ends when either a local optimal solution is found or some other termination condition is met.

\subsection{HT LP.}
The LP problem used for the HT SLP is reffered to as the HT LP for this paper. For the HT LP it is assumed that the network is 3-phase, that reverse power flow is acceptable, that all excess active power can be exported to the upstream grid and will be bought.

The HT LP is shown in (\ref{eq:NPV}) and (\ref{eq:OPF_objfunc})-(\ref{eq:OPF_HTbin}). For the HT LP, (\ref{eq: nonlin_epq})-(\ref{eq: nonlin_Qmax}) have been linearised using First Order Taylor Series approximation, resulting in (\ref{eq:OPF_epq})-(\ref{eq:OPF_Qconv}). Additionally, in order to limit how far controllable loads can deviate from their nomial operating points, step-size limits have been applied to DG power outputs and (\ref{eq:nonlin_r})-(\ref{eq:nonlin_gamma}). $L_{00}$, $L_{01}$, $L_0$, $L_1$, $L_2$, $L_3$, $L_4$, and $L_5$ defined in (\ref{eq:shorthand00})-(\ref{eq:shorthand5}) simply serve as a form of shorthand used in (\ref{eq:OPF_realvolt})-(\ref{eq:OPF_Psub}), (\ref{eq:OPF_epq}), and (\ref{eq:OPF_Ppri})-(\ref{eq:OPF_Qconv}) to improve constraint layout and save space.  For the HT LP, the approximated annual income earned from DG active power exports, $C^\mathrm{exp} \in \mathbb{R}$, is calculated by summing together the power exports made at the slack bus, $P^\mathrm{sub} \in \mathbb{R}^{\Omega}$, for all time periods. How $P^\mathrm{sub}$ values are summed together and adjusted depends on the daily load profiles used to approximate the year. $P^\mathrm{sub}$ summations are converted from per unit to kW and then multiplied by the energy export price, $c^\mathrm{e} \in \mathbb{R}$, to calculate $C^\mathrm{exp}$.

\patchcmd\subequations
 {\theparentequation\alph{equation}}
 {\subequationsformat}
 {}{}
\newcommand{\subequationsformat}{\theparentequation.\arabic{equation}}

\begingroup
\allowdisplaybreaks
\begin{align}
    NPV &= \begin{aligned}[t]&\sum_{tp \in HT^\mathrm{yrs}}\frac{C^\mathrm{exp}-C^\mathrm{ref}}{\left(1+coc\right)^{tp}} - \sum_{ij \in \mathcal{N}^\mathrm{HT}}b_{ij}^\mathrm{HT}C_{ij}^\mathrm{inv}
    \end{aligned} \label{eq:NPV}
\end{align}
\endgroup

\begingroup
\allowdisplaybreaks
\begin{subequations}
\begin{equation}
    \max_{} \ \begin{aligned}[t]
         &NPV - \sum_{t \in \mathcal{N}^\mathrm{t}}\sum_{ij \in \varepsilon^\mathrm{HT}}\sum_{\phi \in \Omega}\bigl(w_1(e_{t,ij,\phi}^\mathrm{abs}+f_{t,ij,\phi}^\mathrm{abs}) \\
         &- w_2Q_{t,ij,\phi}^\mathrm{abs} - w_3 \gamma_{t,ij,\phi}^\mathrm{abs}\bigr)
    \end{aligned} \label{eq:OPF_objfunc} \\
\end{equation}
\begin{align}
    \text{s.t.} & \nonumber \\
    e_{t} &= \begin{aligned}[t]
      &Re\bigl(M^0_{t}+M^\mathrm{Y}_{t}L_{00}s^\mathrm{base}\\&+M^\Delta_{t}L_{01}s^\mathrm{base} \bigr)/v^\mathrm{base},
       \end{aligned} \label{eq:OPF_realvolt}\\
    f_{t} &=\begin{aligned}[t]
        &Im\bigl(M^0_{t}+M^\mathrm{Y}_{t}L_{00}s^\mathrm{base}\\ &+M^\Delta_{t}L_{01}s^\mathrm{base}\bigr)/v^\mathrm{base},
        \end{aligned} \label{eq:OPF_imagvolt}\\
    \lvert v_{t} \rvert &= \begin{aligned}[t]
        &\bigl(K^0_{t}+K^\mathrm{Y}_{t}L_{00}s^\mathrm{base}\\&+K^\Delta_{t}L_{01}s^\mathrm{base}\bigr)/v^\mathrm{base},
    \end{aligned} \label{eq:OPF_voltmag}\\
    P^\mathrm{sub}_{t} &= \begin{aligned}[t]
      &Re\bigl(A^0_{t}+A^\mathrm{Y}_{t}L_{00}s^\mathrm{base}\\ &+A^\Delta_{t}L_{01}s^\mathrm{base} \bigr)/s^\mathrm{base},
       \end{aligned} \label{eq:OPF_Psub}\\
    \lvert \underline{v} \rvert &\leq \lvert v_{t} \rvert \leq \lvert \overline{v} \rvert, \label{eq:OPF_voltlims}\\
    \Delta x^\Delta_{t} &= ((\Delta p^\Delta_{t})^T,(\Delta q^\Delta_{t})^T)^T, \label{eq:OPF_xdel}\\
    \Delta x^\mathrm{Y}_{t} &= \begin{aligned}[t]&((\Delta p^\mathrm{Y}_{t})^T,(\Delta q^\mathrm{Y}_{t})^T)^T, \end{aligned}\label{eq:OPF_xwye}\\
    \Delta x^\mathrm{HT}_{t} &= \begin{aligned}[t]&((\Delta p^\mathrm{HT}_{t})^{T},(\Delta q^\mathrm{HT}_{t})^{T})^{T},\ \forall t \in \mathcal{N}^\mathrm{t}. \label{eq:OPF_xHT} \end{aligned}\\
    \Delta p_{t, i,\phi}^\Delta &= \begin{aligned}[t]&-P_{t, i,\phi}^\mathrm{DG}, \ \Delta q_{t, i,\phi}^\Delta = -Q_{t, i,\phi}^\mathrm{DG}, \\ 
    &\forall t \in \mathcal{N}^\mathrm{t}, \ \forall i \in \mathcal{N}^\mathrm{DG}\cap \mathcal{N}^{\Delta},\ \forall \phi \in \Omega.\end{aligned}\label{eq:OPF_pdel_Pdg}\\
    \Delta p_{t, i,\phi}^\Delta &= \begin{aligned}[t]&0, \ \Delta q_{t, i,\phi}^\Delta=0, \\&\forall t \in \mathcal{N}^\mathrm{t}, \ \forall i \in \mathcal{N}^{\Delta}\backslash \mathcal{N}^\mathrm{DG}, \ \forall \phi \in \Omega.\end{aligned} \label{eq:OPF_pdel_zero}\\
    \Delta p_{t, i,\phi}^\mathrm{Y} &= \begin{aligned}[t]&-P_{t, i,\phi}^\mathrm{DG}, \ \Delta q_{t, i,\phi}^\mathrm{Y}=-Q_{t, i,\phi}^\mathrm{DG}, \\& \forall t \in \mathcal{N}^\mathrm{t}, \ \forall i \in \mathcal{N}^\mathrm{DG} \cap \mathcal{N}^\mathrm{Y},\ \forall \phi \in \Omega.\end{aligned} \label{eq:OPF_pwye_Pdg}\\
    \Delta p_{t, i,\phi}^\mathrm{Y} &= \begin{aligned}[t] &0, \ \Delta q_{t, i,\phi}^\mathrm{Y}=0  \\ &\forall t \in \mathcal{N}^\mathrm{t}, \ \forall i \in \mathcal{N}^\mathrm{Y}\backslash \mathcal{N}^\mathrm{DG}, \forall \phi \in \Omega.\end{aligned} \label{eq:OPF_pwye_zero}\\
    P_{t,i,\phi}^\mathrm{DG} &\leq \begin{aligned}[t] &min(\overline{P}_{i},P_{t,i,\phi}^{\mathrm{DG}(k)}+stp_{i,\phi}^\mathrm{DG}),\end{aligned} \label{eq:OPF_Pdg_leq}\\
    P_{t,i,\phi}^\mathrm{DG} &\geq \begin{aligned}[t] &max(0,P_{t,i,\phi}^{\mathrm{DG}(k)}-stp_{i,\phi}^\mathrm{DG}), \end{aligned} \label{eq:OPF_Pdg_geq}\\
    Q_{t,i,\phi}^\mathrm{DG} &= \begin{aligned}[t] &P_{t,i,\phi}^\mathrm{DG} \frac{\sqrt{1-PF_{i}^2}}{PF_{i}},
    \\  &\forall t \in \mathcal{N}^\mathrm{t}, \ \forall i \in \mathcal{N}^\mathrm{DG}, \ \forall \phi \in \Omega.
    \end{aligned} \label{eq:OPF_Qdg}\\
    P_{t,i,0}^\mathrm{DG} &= \begin{aligned}[t]&P_{t,i,1}^\mathrm{DG} = P_{t,i,2}^\mathrm{DG}, \\& \forall t \in \mathcal{N}^\mathrm{t}, \ \forall i \in \mathcal{N}^\mathrm{DG}. \end{aligned} \label{eq:OPF_Pdg_bal}\\
    \Delta p_{t, i, \phi}^\mathrm{HT} &= \begin{aligned}[t] &0, \ \Delta q_{t, i, \phi}^\mathrm{HT} = 0, \ \forall t \in \mathcal{N}^\mathrm{t}, \\ &\forall i \in \mathcal{N}^\mathrm{Y} \backslash (\mathcal{N}^\mathrm{HTp} \cup \mathcal{N}^\mathrm{HTs}), \forall \phi \in \Omega.
    \end{aligned} \label{eq:OPF_pHT_0}\\
    \Delta p_{t,i,\phi}^\mathrm{HT} &= \begin{aligned}[t]&-P_{t,ij,\phi}^\mathrm{pri}, \ \Delta p_{t,j,\phi}^\mathrm{HT}=-P_{t,ij,\phi}^\mathrm{sec},\end{aligned} \label{eq:OPF_pwye_Ppri} \\
    \Delta q_{t,i,\phi}^\mathrm{HT} &= \begin{aligned}[t]&-Q_{t,ij,\phi}^\mathrm{pri}, \ \Delta q_{t,j,\phi}^\mathrm{HT}=-Q_{t,ij,\phi}^\mathrm{sec},\end{aligned} \label{eq:OPF_qwye_Qpri}\\
    r_{t,ij,\phi} &\leq \begin{aligned}[t] &min(\overline{r}, r_{t,ij,\phi}^{(k)}+stp^{v}),
    \end{aligned}\label{eq:OPF_rmag_leq}\\
    r_{t,ij,\phi} &\geq \begin{aligned}[t] &max(\underline{r}, r_{t,ij,\phi}^{(k)}-stp^{v}),\end{aligned} \label{eq:OPF_rmag_geq}\\
    \gamma_{t,ij,\phi} &\leq \begin{aligned}[t] &min(\pi,\gamma_{t,ij,\phi}^{(k)}+stp^{\gamma}),\end{aligned} \label{eq:OPF_gamma_leq}\\
    \gamma_{t,ij,\phi} &\geq \begin{aligned}[t] &max(-\pi,\gamma_{t,ij,\phi}^{(k)}-stp^{\gamma}), \end{aligned} \label{eq:OPF_gamma_qeg}\\
    \gamma_{t,ij,\phi}^\mathrm{abs} &\geq \gamma_{t,ij,\phi}, \ \gamma_{t,ij,\phi}^\mathrm{abs} \geq -\gamma_{t,ij,\phi}, \label{eq:OPF_gamma_abs1}\\
    -\lvert \overline{Q}_{t,ij,\phi}\rvert &\leq \begin{aligned}[t]&Q_{t,ij,\phi}^\mathrm{sh} \leq \lvert \overline{Q}_{t,ij,\phi}\rvert,
    \end{aligned} \label{eq:OPF_Qsh}\\
    Q_{t,ij,\phi}^\mathrm{abs} &\geq \begin{aligned}[t]&Q_{t,ij,\phi}^\mathrm{sh}, \  Q_{t,ij,\phi}^\mathrm{abs} \geq-Q_{t,ij,\phi}^\mathrm{sh},
    \end{aligned}\label{eq:OPF_Qabs_pos}\\
    -\lvert \overline{Q}_{t,ij,\phi}\rvert &\leq \begin{aligned}[t]&Q_{t,ij,\phi}^\mathrm{c} \leq \lvert \overline{Q}_{t,ij,\phi}\rvert,
    \end{aligned} \label{eq:OPF_Qconv_lims}\\
    P_{t,ij,\phi}^\mathrm{sec} &= \begin{aligned}[t]&P_{t,ij,\phi}^\mathrm{se}-\left(P_{t,ij,\phi}^\mathrm{sh}/md^\mathrm{HT}\right),
    \end{aligned} \label{eq:OPF_Psec}\\
    -\overline{S}_{t,ij,\phi} &\leq \begin{aligned}[t]&P_{t,ij,\phi}^\mathrm{sh} \leq  \overline{S}_{t,ij,\phi},
    \end{aligned} \label{eq:OPF_Psh_lims}\\
    Q_{t,ij,\phi}^\mathrm{sec} &= \begin{aligned}[t]&Q_{t,ij,\phi}^\mathrm{se}+\left(Q_{t,ij,\phi}^\mathrm{sh}/md^\mathrm{HT}\right),
    \end{aligned} \label{eq:OPF_Qsec}\\
    e_{t,ij,\phi}^\mathrm{abs} &\geq \begin{aligned}[t] &e_{t,ij,\phi}^\mathrm{pq}, \ e_{t,ij,\phi}^\mathrm{abs} \geq -e_{t,ij,\phi}^\mathrm{pq}\end{aligned} \label{eq:OPF_eabs_pos}\\
    f_{t,ij,\phi}^\mathrm{abs} &\geq \begin{aligned}[t]&f_{t,ij,\phi}^\mathrm{pq}, \ f_{t,ij,\phi}^\mathrm{abs} \geq -f_{t,ij,\phi}^\mathrm{pq},\end{aligned} \label{eq:OPF_fabs_pos}\\
    e_{t,ij,\phi}^\mathrm{pq} &=\begin{aligned}[t] &Re\bigl(L_0\bigr), \ f_{t,ij,\phi}^\mathrm{pq} = Im\bigl(L_0\bigr),\end{aligned} \label{eq:OPF_epq}\\
    \lvert \overline{Q}_{t,ij,\phi}\rvert &= \begin{aligned}[t]
    &C_1(((\overline{S}_{ij,\phi})^2 -(P_{t,ij,\phi}^{\mathrm{sh}(k)})^2)^{\frac{1}{2}}\\
     &-(P_{t,ij,\phi}^{\mathrm{sh}(k)}((\overline{S}_{ij,\phi})^2 \\
     &-(P_{t,ij,\phi}^{\mathrm{sh}(k)})^2)^{-\frac{1}{2}}(P_{t,ij,\phi}^\mathrm{sh}-P_{t,ij,\phi}^{\mathrm{sh}(k)}))),\end{aligned} \label{eq:OPF_Qmax}\\
    P_{t,ij,\phi}^\mathrm{pri}&= \begin{aligned}[t] 
    &\bigl(-G_{ij,\phi\phi}L_1+B_{ij,\phi\phi}L_2\bigr)/md^\mathrm{HT},
    \end{aligned} \label{eq:OPF_Ppri}\\
    Q_{t,ij,\phi}^\mathrm{pri} &= \begin{aligned}[t]
    &\bigl(G_{ij,\phi\phi}L_2+B_{ij,\phi\phi}L_1\bigr)/md^\mathrm{HT},
    \end{aligned} \label{eq:OPF_Qpri}\\
    P_{t,ij,\phi}^\mathrm{se}&= \begin{aligned}[t]
    &\bigl(G_{ij,\phi\phi}L_3+B_{ij,\phi\phi}L_4\bigr)/md^\mathrm{HT},
    \end{aligned} \label{eq:OPF_Pse}\\
    Q_{t,ij,\phi}^\mathrm{se}&= \begin{aligned}[t]
    &\bigl(G_{ij,\phi\phi}L_4-B_{ij,\phi\phi}L_3\bigr)/md^\mathrm{HT},
    \end{aligned} \label{eq:OPF_Qse}\\
    P_{t,ij,\phi}^\mathrm{sh} &= \begin{aligned}[t] &G_{ij,\phi\phi}\bigl(L_5+L_1-L_3\bigr)\\&+B_{ij,\phi\phi}\bigl(L_2+L_4\bigr),
    \end{aligned} \label{eq:OPF_Psh}\\
    Q_{t,ij,\phi}^\mathrm{c}&=\begin{aligned}[t]&G_{ij,\phi\phi}\bigl(L_2+L_4\bigr)\\
    &-B_{ij,\phi\phi}\bigl(L_5+L_1-L_3\bigr),\\ 
    &\forall t \in \mathcal{N}^\mathrm{t},\ \forall(i,j) \in \varepsilon^\mathrm{HT},\ \forall \phi \in \Omega.
    \end{aligned} 
    \label{eq:OPF_Qconv}\\
    b_{ij}^\mathrm{HT} &\geq \begin{aligned}[t]&C_2\sum_{t \in \mathcal{N}^\mathrm{t}}\sum_{\phi \in \Omega}\left(e_{t, ij, \phi}^\mathrm{abs}+f_{t, ij, \phi}^\mathrm{abs}\right.
    \\& \left. +\gamma_{t, ij, \phi}^\mathrm{abs} +Q_{t, ij, \phi}^\mathrm{abs}\right),\ \forall(i,j) \in \varepsilon^\mathrm{HT}.
    \end{aligned}
    \label{eq:OPF_HTbin}
\end{align}
\end{subequations}
\endgroup

\begingroup
\allowdisplaybreaks
\begin{align} 
L_{00}&= x^\mathrm{Y}_{t}+\Delta x^\mathrm{Y}_{t} + \Delta x^\mathrm{HT}_{t}md^\mathrm{HT}, L_{01} = x^{\Delta}_{t}+\Delta x^{\Delta}_{t}, \label{eq:shorthand00}\\
L_0 &=\begin{aligned}[t] &r_{t,ij,\phi}^{(k)}v_{t,i,\phi}^{(k)}exp(-j\gamma_{t,ij,\phi}^{(k)}) \\
     &+v_{t,i,\phi}^{(k)}exp(-j\gamma_{t,ij,\phi}^{(k)})(r_{t,ij,\phi}-r_{t,ij,\phi}^{(k)})\\
     &+ r_{t,ij,\phi}^{(k)}exp(-j\gamma_{t,ij,\phi}^{(k)})((e_{t,i,\phi}+jf_{t,i,\phi})-v_{t,i,\phi}^{(k)})\\ 
     &-jr_{t,ij,\phi}^{(k)}v_{t,i,\phi}^{(k)}exp(-j\gamma_{t,ij,\phi}^{(k)})(\gamma_{t,ij,\phi}-\gamma_{t,ij,\phi}^{(k)}),
\end{aligned}\label{eq:shorthand0}\\
L_1 &=\begin{aligned}[t]&\left(e_{t,ij,\phi}^{\mathrm{pq}(k)}e_{t,i,\phi}+e_{t,i,\phi}^{(k)}e_{t,ij,\phi}^\mathrm{pq}-e_{t,i,\phi}^{(k)}e_{t,ij,\phi}^{\mathrm{pq}(k)}\right)\\&+\left(f_{t,ij,\phi}^{\mathrm{pq}(k)}f_{t,i,\phi}+ f_{t,i,\phi}^{(k)}f_{t,ij,\phi}^\mathrm{pq}-f_{t,i,\phi}^{(k)}f_{t,ij,\phi}^{\mathrm{pq}(k)}\right),
\end{aligned} \label{eq:shorthand1}\\
L_2 &=\begin{aligned}[t]&\left(f_{t,ij,\phi}^{\mathrm{pq}(k)}e_{t,i,\phi}+e_{t,i,\phi}^{(k)}f_{t,ij,\phi}^\mathrm{pq}-e_{t,i,\phi}^{(k)}f_{t,ij,\phi}^{\mathrm{pq}(k)}\right)\\&-\left(e_{t,ij,\phi}^{\mathrm{pq}(k)}f_{t,i,\phi}+ f_{t,i,\phi}^{(k)}e_{t,ij,\phi}^\mathrm{pq}-f_{t,i,\phi}^{(k)}e_{t,ij,\phi}^{\mathrm{pq}(k)}\right),
\end{aligned} \label{eq:shorthand2}\\
L_3 &=\begin{aligned}[t]&\left(e_{t,ij,\phi}^{\mathrm{pq}(k)}e_{t,j,\phi}+e_{t,j,\phi}^{(k)}e_{t,ij,\phi}^\mathrm{pq}-e_{t,j,\phi}^{(k)}e_{t,ij,\phi}^{\mathrm{pq}(k)}\right)\\&+\left(f_{t,ij,\phi}^{\mathrm{pq}(k)}f_{t,j,\phi}+ f_{t,j,\phi}^{(k)}f_{t,ij,\phi}^\mathrm{pq}-f_{t,j,\phi}^{(k)}f_{t,ij,\phi}^{\mathrm{pq}(k)}\right),
\end{aligned} \label{eq:shorthand3}\\
L_4 &=\begin{aligned}[t]&\left(e_{t,ij,\phi}^{\mathrm{pq}(k)}f_{t,j,\phi}+ f_{t,j,\phi}^{(k)}e_{t,ij,\phi}^\mathrm{pq}-f_{t,j,\phi}^{(k)}e_{t,ij,\phi}^{\mathrm{pq}(k)}\right)\\&-
\left(f_{t,ij,\phi}^{\mathrm{pq}(k)}e_{t,j,\phi}+e_{t,j,\phi}^{(k)}f_{t,ij,\phi}^\mathrm{pq}-e_{t,j,\phi}^{(k)}f_{t,ij,\phi}^{\mathrm{pq}(k)}\right),
\end{aligned} \label{eq:shorthand4}\\
L_5 &=\begin{aligned}[t]& -\left(e_{t,ij,\phi}^{\mathrm{pq}(k)}\right)^2-\left( f_{t,ij,\phi}^{\mathrm{pq}(k)} \right)^2+2e_{t,ij,\phi}^\mathrm{pq}e_{t,ij,\phi}^{\mathrm{pq}(k)}\\&+2f_{t,ij,\phi}^\mathrm{pq}f_{t,ij,\phi}^{\mathrm{pq}(k)}.
\end{aligned} \label{eq:shorthand5}
\end{align}
\endgroup
where notation $a_t$ indicates to which time period decision variables and constants belong to (e.g., $e^\mathrm{pq}_{t,ij,\phi}$ is the $e^\mathrm{pq}_{ij,\phi}$  value for time period $t$). $A^0 \in \mathbb{C}^{\Omega}$ is the linear network model zero-load slack bus power flow vector, and $A^\mathrm{Y}, A^{\Delta} \in \mathbb{C}^{\Omega \times 6\mathcal{N}}$ are the linear network model coefficient matrices for power injections made at wye-connected and delta-connected buses to calculate slack bus power flow \cite{LinBernstein2018, LinBernstein2017}. $b_{ij}^\mathrm{HT} \in \mathbb{R}^{\mathcal{N}^\mathrm{HT}}$ is a binary array used to identify which HTs are installed. $C_{1} \in \mathbb{R}$ is a constant used to reduce results of (\ref{eq:OPF_Qmax}) to ensure feasible results, and $C_{2} \in \mathbb{R}$ is a constant used to reduce the value of the summation of HT compensation to be in the range between 0 and 1 in (\ref{eq:OPF_HTbin}). $C^\mathrm{inv} \in \mathbb{R}^{\mathcal{N}^\mathrm{HT}}$ is the array of HT investment costs. $C^\mathrm{ref} \in \mathbb{R}$ is the maximum possible annual income earned without HT utilisation. $coc \in \mathbb{R}$ is the discount rate. $e, f \in  \mathbb{R}^{3\mathcal{N}}$ are the real and imaginary network phase voltages. $e^\mathrm{abs}_{ij}, f^\mathrm{abs}_{ij} \in \mathbb{R}^{\Omega}$ are the magnitudes of of $e^\mathrm{pq}_{ij}$ and $f^\mathrm{pq}_{ij}$. $HT^\mathrm{yrs} \in \mathbb{R}$ is the number of years in the HTs' operational lifespan. $K^0 \in \mathbb{R}^{3\mathbb{N}}$ is the linear network model zero-load voltage magnitude vector, and $K^\mathrm{Y}, K^{\Delta} \in \mathbb{R}^{3\mathcal{N} \times 6\mathcal{N}}$ are the linear network model coefficient matrices for power injections made at wye-connected and delta-connected buses to calculate voltage magnitudes \cite{LinBernstein2018, LinBernstein2017}. $M^0 \in \mathbb{C}^{3\mathcal{N}}$ is the linear network model zero-load complex voltage vector, and $M^\mathrm{Y}, M^{\Delta} \in \mathbb{C}^{3\mathcal{N} \times 6\mathcal{N}}$ are the linear network model coefficient matrix for power injections made at wye-connected and delta-connected buses to calculate complex voltages \cite{LinBernstein2018, LinBernstein2017}. $md^\mathrm{HT} \in \mathbb{R}$ is a constant used to adjust scale of HT power injections to prevent numerical instability. $\overline{P_i} \in \mathbb{R}$ is the maximum allowable active power output for DG connected at network bus $i$. $P^\mathrm{DG}_{i}, Q^\mathrm{DG}_{i} \in \mathbb{R}^{\Omega}$ are the active and reactive power outputs of DG connected at network bus $i$. $PF_i \in \mathbb{R}$ is the power factor for DG connected at network bus $i$. $P^\mathrm{sub} \in \mathbb{R}^{\Omega}$ is the active power exported/imported at the slack bus. $Q^\mathrm{abs}_{t,ij} \in \mathbb{R}^{\Omega}$ is the magnitude of $Q^\mathrm{sh}_{ij}$. $s^\mathrm{base} \in \mathbb{R}$ is the network base apparent power. $stp^\mathrm{DG}_{i} \in \mathbb{R}$ is the step-size restriction for DG active power output at network bus $i$, and $\underline{stp}^\mathrm{DG}_{i} \in \mathbb{R}$ is the minimum allowable step-size restriction for DG active power output at network bus $i$. $stp^{v}, stp^{\gamma} \in \mathbb{R}$ are the step-size restrictions for HT voltage compensation and phase-shifting. $\lvert v \rvert \in \mathbb{R}^{3\mathcal{N}}$ is the network phase voltage magnitudes, and $\lvert \underline{v} \rvert, \lvert \overline{v} \rvert \in \mathbb{R}$ are the minimum and maximum allowable network phase voltages. $v^\mathrm{base} \in \mathbb{R}$ is the network base voltage. $x^\mathrm{Y}, x^{\Delta} \in \mathbb{R}^{6\mathcal{N}}$ are uncontrollable network loads at wye-connected and delta-connected buses. $\Delta p^\mathrm{HT}, \Delta q^\mathrm{HT} \in \mathbb{R}^{3\mathcal{N}}$ are the network active and reactive power injections due to HTs. $\Delta p^\mathrm{Y}, \Delta q^\mathrm{Y},\Delta p^{\Delta}, \Delta q^{\Delta} \in \mathbb{R}^{3\mathcal{N}}$ are the controllable network active and reactive power injections at wye-connected and delta-connected buses. $\Delta x^\mathrm{HT} \in \mathbb{R}^{6\mathcal{N}}$ are the controllable network power injections due to HTs. $\Delta x^{Y}, \Delta x^{\Delta} \in \mathbb{R}^{6\mathcal{N}}$ are the controllable network power injections at wye-connected and delta-connected buses due to DGs.

The objective function of the HT LP is to maximise the NPV possible from exporting excess DG active power over the operational lifespan of the HTs while also minimising the HT compensation provided with the weights $w_1, w_2, w_3 \in \mathbb{R}$ (which determine to what extent HT utilisation is discouraged) and variables $e^\mathrm{abs}_{ij}$, $f^\mathrm{abs}_{ij}$, $Q^\mathrm{abs}_{ij}$, and $\gamma^\mathrm{abs}_{ij}$ (which track the magnitudes of HT compensation) as shown in (\ref{eq:OPF_objfunc}). HT compensation is discouraged to prevent HTs from providing more compensation than is required to prevent voltage violations, which may result in failed HT accuracy checks, which can lead to the HT SLP prematurely terminating (as shown later in this section). Several factors affect what values to use for the weights, such as the NPV, network voltage levels, HT power ratings, HT investment costs, HT operational lifespan, and the number of time periods. As such, the values for weights will depend heavily on the test case. However, in general, the values for weights should scale the summations of HT compensation such that they are around 1 to 3 orders of magnitude smaller than the NPV calculated for all HT SLP iterations where HTs are utilised.

As previosuly mentioned, in (\ref{eq:OPF_Ppri})-(\ref{eq:OPF_Qse}) the final values are divided by $md^\mathrm{HT}$ to prevent numeric instability. This is because $P^\mathrm{pri}$, $Q^\mathrm{pri}$, $P^\mathrm{se}$, and $Q^\mathrm{se}$ are much larger than other power injections, such that numeric instability can occur. $\Delta x^\mathrm{HT}$ is then multiplied by $md^\mathrm{HT}$ in (\ref{eq:shorthand00}) to restore the power injections to their true values.

For the HT LP, the following terms are constants: $A^0$, $A^\mathrm{Y}$, $A^{\Delta}$, $B_{ij}$, $C_{1}$, $C_{2}$, $C^\mathrm{inv}$, $C^\mathrm{ref}$,$c^\mathrm{e}$, $coc$, $G_{ij}$, $K^0$, $K^\mathrm{Y}$, $K^{\Delta}$, $M^0$, $M^\mathrm{Y}$, $M^{\Delta}$, $md^\mathrm{HT}$, $\overline{P}_{i}$, $PF_{i}$, $s^\mathrm{base}$, $stp^\mathrm{DG}$, $stp^{v}$, $stp^{\gamma}$, $\lvert\underline{v}\rvert$, $\lvert\overline{v}\rvert$, $v^\mathrm{base}$, $x^\mathrm{Y}$, $x^{\Delta}$. The notation $a^{(k)}$ is used to denote nominal operating points constants for the HT LP in iteration $k$ of the HT SLP. All other components of the HT LP, excluding sets, are decision variables.

\subsection{HT SLP Process.}
\begin{figure}[t!]
    \centering
    \includegraphics[width=0.75\linewidth]{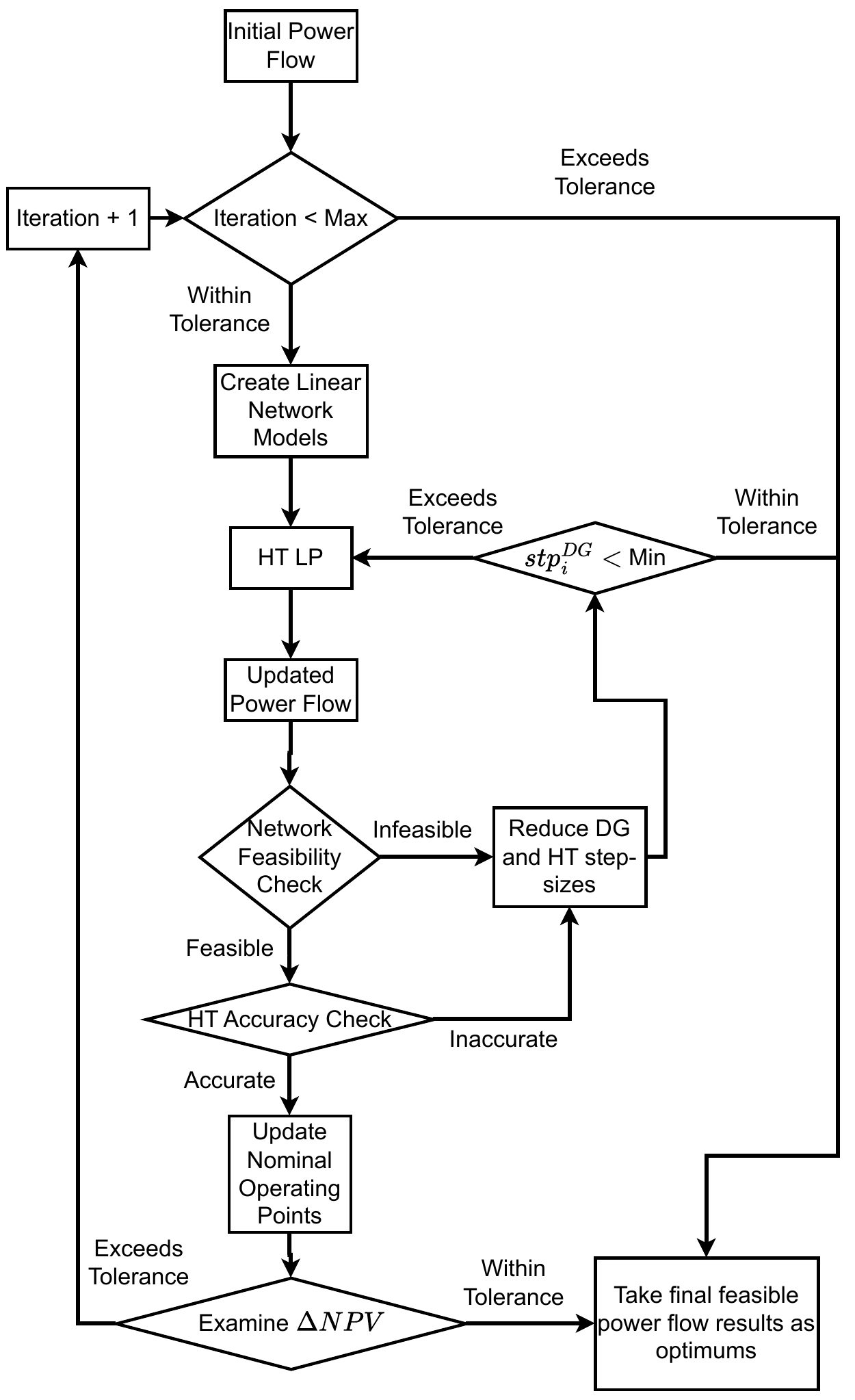}
    \caption{HT SLP Flowchart.}
    \label{fig:HT SLP Flowchart}
\end{figure}

The HT SLP method is a process of solving the HT LP, confirming whether HT LP results are feasible and accurate, adjusting step-sizes in response to any feasibility or accuracy issues, updating HT LP inputs, and determining whether a local optimal solution has been reached, as shown in Figure \ref{fig:HT SLP Flowchart}.

Every HT SLP iteration begins with a check to ensure that the current iteration number, $k$, does not exceed the maximum allowable number of iterations, $\overline{k}$. If $k\leq\overline{k}$, then the HT SLP continues onto the next step, otherwise, the HT SLP will end, with the HT LP results from the previous iteration, $k-1$, being taken as the final result.

For the first HT SLP iteration, the HT LP inputs are collected from a series of Z-bus power simulations \cite{Bazrafshan2018, Bazrafshan20182}, one for every time period, where all HTs and DGs are inactive. As such, $e^{\mathrm{pq} \ (k)}$, $f^{\mathrm{pq} \ (k)}$, $r^{(k)}$, $\gamma^{(k)}$, $P^{\mathrm{sh}(k)}$, and $P^{DG(k)}$ are set to zero, while the values for the linear network matrices and the nominal operating points $v^{(k)}$, $e^{(k)}$, and $f^{(k)}$ are derived from the power flow results. After HT LP results successfully pass the feasibility and accuracy checks, they are used to update the nominal operating points for the next HT SLP iteration. Updated linear network model matrices are derived at the start of the next iteration using the updated power flow results of the previous iteration. These updated nominal operating points and linear network model matrices serve as the inputs for the next HT LP.

There are two main checks in the HT SLP to determine if HT LP results are feasible and accurate. The first is the `Network Feasibility Check', which checks for any significant voltage violations in the updated power flow results for all of the time periods. If there are any significant voltage violations in any of the time periods, then both DG and HT step-sizes are reduced by 50\%, and the HT LP is reran. For the `Network Feasibility Check' it is recommended that the magnitude of network voltage levels be rounded to some extent, as to avoid minor violations (e.g., less than 0.005pu) from requiring the HT LP to be reran. The second check is the `HT Accuracy Check' which determines whether the HT compensation predicted by the HT LP is accurate enough to the HT compensation actually obtained. In the `HT Accuracy Check', the actual series converter apparent power and compensating voltages are calculated using the results from the HT LP and the updated power flow simulations, as shown in (\ref{eq:vpq1})-(\ref{eq:Sconv_calc}).

\begingroup
\allowdisplaybreaks
\begin{align}
v_{t,ij,\phi}^{\mathrm{pq}1} &= \begin{aligned}[t]&-Z_{ij, \phi\phi}^\mathrm{HT}\left(\frac{S_{t,ij, \phi}^\mathrm{pri}}{v_{t,i,\phi}}\right)^*,\end{aligned} \label{eq:vpq1}\\
v_{t,ij,\phi}^{\mathrm{pq}2} &= \begin{aligned}[t]&Z_{ij, \phi\phi}^\mathrm{HT}\left(\frac{S_{t,ij, \phi}^\mathrm{se}}{v_{t,j,\phi}}\right)^*,\end{aligned} \label{eq:vpq2}\\
v_{t,ij,\phi}^\mathrm{pq\ avg} &= \begin{aligned}[t]&\frac{v_{t,ij,\phi}^{\mathrm{pq}1}+v_{t,ij,\phi}^{\mathrm{pq}2}}{2},\end{aligned}\label{eq:vpqact}\\
\Delta v_{t,ij,\phi}^\mathrm{pq} &= \begin{aligned}[t]&\frac{\lvert v_{t,ij,\phi}^\mathrm{pq \ avg} - v_{t,ij,\phi}^\mathrm{pq \ LP}\rvert} {v_{t,ij,\phi}^\mathrm{pq \ LP}}\times100\%,\end{aligned} \label{eq:vpq_delta}\\
S^\mathrm{c}_{t,ij,\phi} &= \begin{aligned}[t]&\biggl\lvert v_{t,ij,\phi}^\mathrm{pq\ avg}\left(\frac{v_{t,i,\phi}+v_{t,ij,\phi}^\mathrm{pq\ avg}-v_{t,j,\phi}}{Z_{ij, \phi\phi}^\mathrm{HT}}\right)^*\biggr\rvert, \\&\forall t \in \mathcal{N}^\mathrm{t}, \forall(i,j) \in \varepsilon^\mathrm{HT},\ \forall \phi \in \Omega.\end{aligned} \label{eq:Sconv_calc}
\end{align}
\endgroup
where $v_{t,ij}^{\mathrm{pq}1}$, $v_{t,ij}^{\mathrm{pq}2} \in \mathbb{C}^{\Omega}$ are the HT compensating voltage calculated from power flow results, $v_{t,ij}^\mathrm{pq\ avg} \in \mathbb{C}^{\Omega}$ is the averaged value of $v_{t,ij}^{\mathrm{pq}1}$ and $v_{t,ij}^{\mathrm{pq}2}$, $v_{t,ij}^\mathrm{pq\ LP} \in \mathbb{C}^{\Omega}$ is the HT compensating voltage predicted by the HT LP, and $\Delta v^\mathrm{pq}_{t,ij} \in \mathbb{R}^{\Omega}$ is the percentage difference in magnitude between $v_{t,ij}^\mathrm{pq\ avg}$ and $v_{t,ij}^\mathrm{pq\ LP}$. Once calculated, all $\Delta v_{t,ij,\phi}^\mathrm{pq}$ values are compared to the predetermined tolerance $v^\mathrm{pq\ tol} \in \mathbb{R}$, and all $S_{t,ij,\phi}^\mathrm{c}$ values are compared to the appropriate $\overline{S}_{ij}$ values. HT LP results are considered inaccurate and step-sizes are reduced by 50\% if any $\Delta v_{t,ij,\phi}^{pq}$ value exceeds $v^\mathrm{pq\ tol}$, or if any $S_{t,ij,\phi}^\mathrm{c}$ value exceeds its appropriate $\overline{S}_{ij}$ value. Otherwise, the HT LP results are considered to be accurate.

Whenever DG step-sizes are reduced, they are compared to a set minimum tolerances, $\underline{stp}_{i}^\mathrm{DG}$. If $stp^\mathrm{DG}_{i}$ is less than $\underline{stp}^\mathrm{DG}_{i}$, then the HT SLP will terminate and the results from the previous feasible iteration are taken as the final results.

Each HT SLP iteration ends with a comparison between the NPV of the current iteration ($k$) and the NPV of the previous iteration ($k-1$), as shown in (\ref{eq:NPV_diff}). If $\Delta NPV$ is less than the tolerance $Obj^\mathrm{tol} \in \mathbb{R}$, then it is considered that the HT SLP has found a local optimal solution and, therefore, terminates. Otherwise, the HT SLP continues onto the next iteration, $k+1$.

\begin{align}
\Delta NPV = \frac{NPV^{(k)}-NPV^{(k-1)}}{NPV^{(k-1)}} \times 100\%
\label{eq:NPV_diff}
\end{align}

\section{RESULTS.}
\label{section:Results}
This section describes the test case network used to determine the effectiveness of the proposed HT SLP, and presents the case study results, demonstrating that the HT SLP is capable of determining the optimal placement and coordination of HTs to generate a positive NPV.

Test cases were run on an HP EliteBook with an Intel i7-10610u, and 32 GB of RAM. Gurobi was used as the solver for the HT LP, and the Z-bus power flow simulations were performed using OPEN\cite{Morstyn2020}.

\subsection{Test Case Setup}
\begin{figure}[t!]
    \centering
    \includegraphics[width=0.6\linewidth]{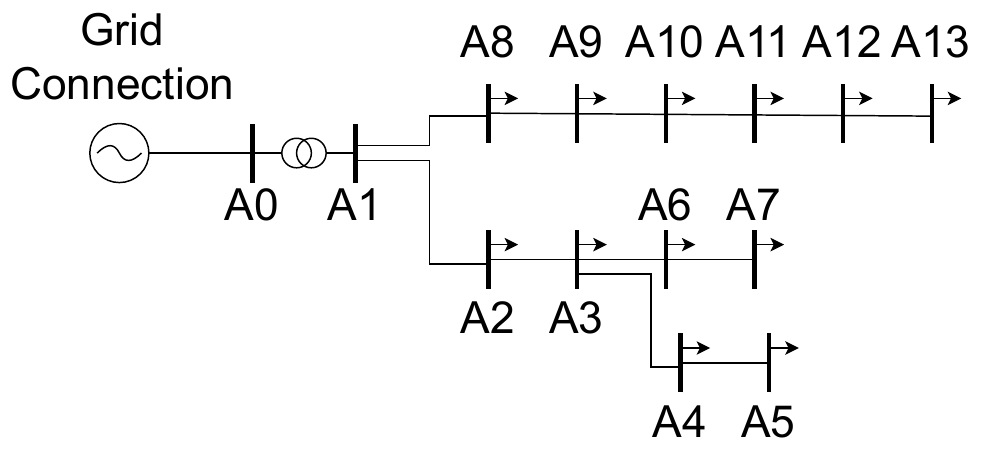}
    \caption{Single line diagram of Benchmark A network}
    \label{fig:CIGRE_and_A}
\end{figure}

\begin{figure}[t!]
    \centering
    \includegraphics[width=0.8\linewidth]{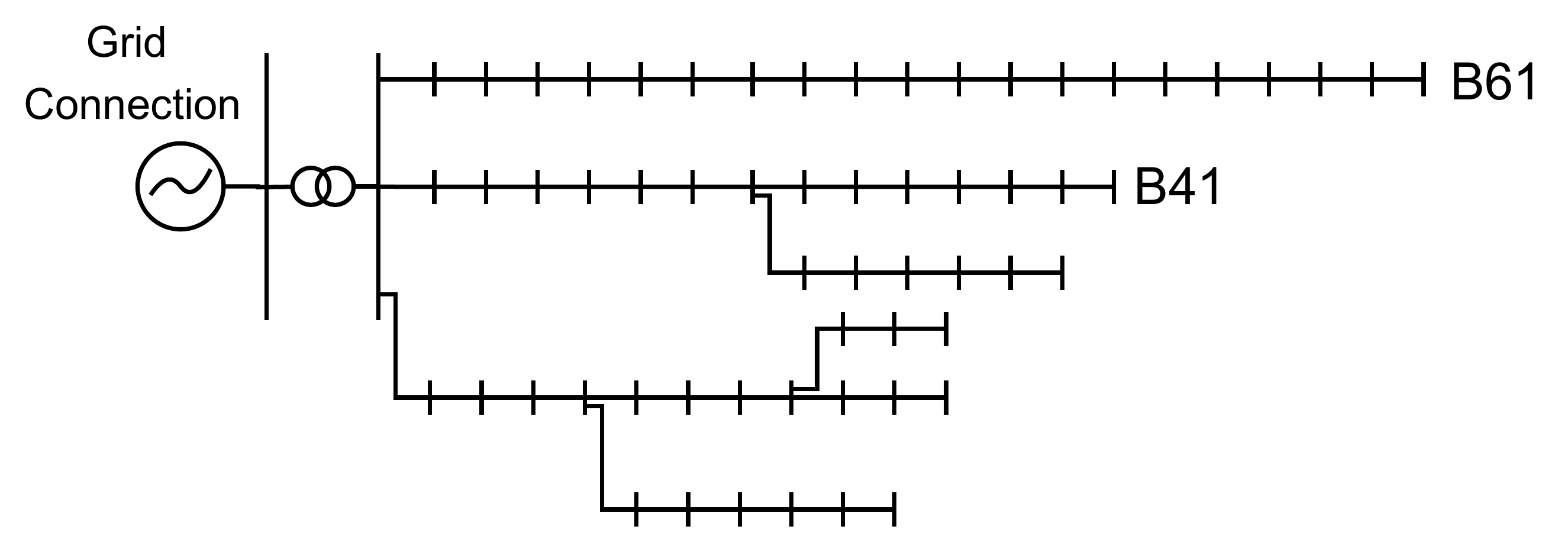}
    \caption{Single line diagram of the Benchmark B network.}
    \label{fig:BenchmarkB}
\end{figure}

\begin{figure}[t!]
    \centering
    \includegraphics[width=0.9\linewidth]{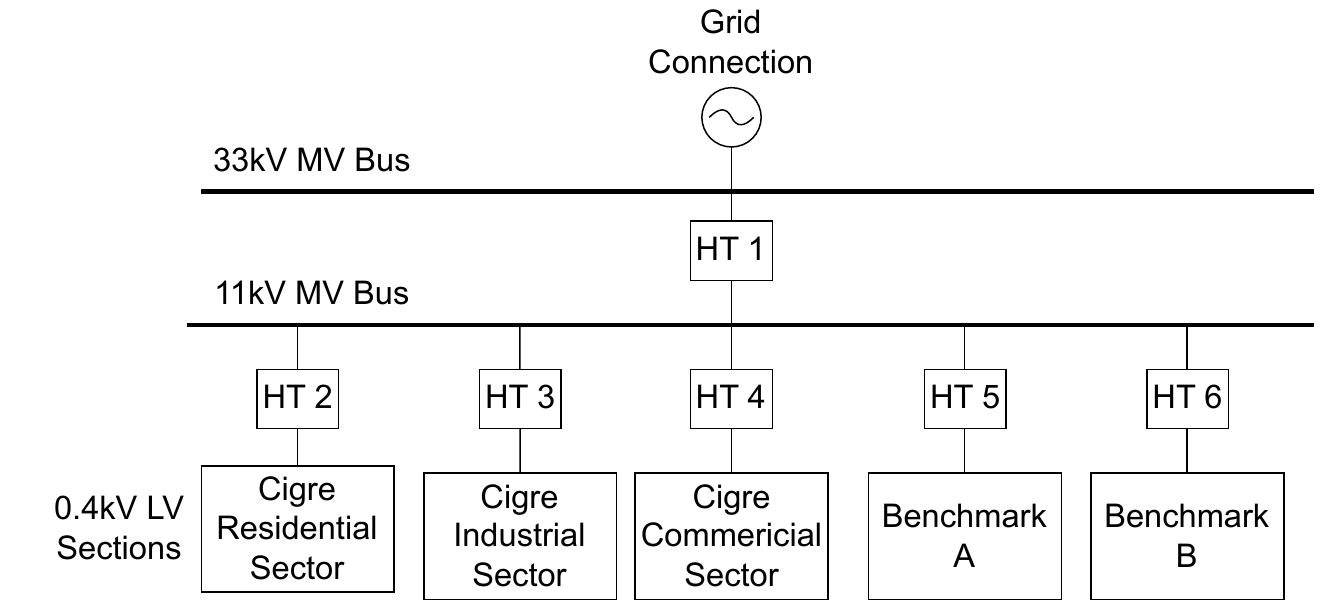}
    \caption{Single line diagram of the test case network.}
    \label{fig:TestCase}
\end{figure}

The test case network used for this paper is a combination of the Cigre European LV distribution network benchmark \cite{CIGRE2014} and 2 other LV network benchmarks (referred to as Benchmarks A and B for this paper, shown in Figures \ref{fig:CIGRE_and_A} and \ref{fig:BenchmarkB}) \cite{Dickert2013, Biener2016}. The combined benchmark test case network is shown in Figure \ref{fig:TestCase}. A $\pm$10\% voltage limit (i.e., $\underline{v} = 0.9$pu and $\overline{v} = 1.1 $pu) was assumed.

For this test case, the HT models were capable of reactive power compensation and $\pm$10\% independent phase voltage regulation ($\underline{r} = -0.1$ and $\overline{r}=0.1$). The power rating of the HT converters, $\overline{S}$, was 1/10$^{th}$ of the transformer's power rating, $\overline{S}^\mathrm{tf} \in \mathbb{R}$. Furthermore, for all HT models, $Z^\mathrm{HT}$ was 1/10$^{th}$ of the transformer's equivalent impedance (based on the HT designed by the company IONATE), and was purely reactive (HT converter power losses were assumed to be negligible).

Modifications have been made to all network benchmarks for these test cases. This includes the upstream 20kV medium voltage (MV) bus in the Cigre benchmark being converted to 11kV. Additionally, the transmission lines in Benchmarks A and B have been converted from NAYY 4x150mm\textsuperscript{2} cables to NA2XY 4x150mm\textsuperscript{2}. DGs have also been added to the test case network, the parameters for which are presented in Table \ref{table:DG}. All DGs operate at unity power factor, only outputting active power. A 33kV bus and 33/11kV 3MVA 3-phase transformer was added to the test case network, as shown in Figure \ref{fig:TestCase}. All LV transformers have had their voltage ratings changed to 11/0.4 kV, and updated transformer power ratings and HT parameters are presented in Table \ref{table:HT Parameters}. For Table \ref{table:HT Parameters}, $R^\mathrm{tf}, X^\mathrm{tf} \in \mathbb{R}$ are the equivalent resistance and reactance values for the three-phase transformers, and $R^\mathrm{HT}, X^\mathrm{HT} \in \mathbb{R}$ are the resistance and reactance values for the self-impedances in $Z^\mathrm{HT}_{ij}$. The costs for the HTs and conventional transformers are shown in Table \ref{table:HT costs}. The costs of the conventional transformers are based on the costs for the Wilson Power Solutions Tier 2 distribution power transformers \cite{WilsonPOwer2025}. It is assumed that HTs are 50.5\% more expensive than conventional transformers, based on the material costs presented in \cite{Burkard2015}.

\begin{table}[t!]
    \centering
    \caption{DG Parameters}
    \scalebox{0.9}{
        \begin{tabular}{|c|c|c|c|c|c|c|c|c|}
            \hline
            \rowcolor{black!20}& DG1 & DG2 & DG3 & DG4 & DG5 & DG6 & DG7 & DG8\\
            \hline
            \cellcolor{black!10}Bus & R18 & I2 & C12 & C20 & A7 & A13 & B41 & B61\\
            \hline
            \cellcolor{black!10}$\overline{P}$ (kW) & 140 & 165 & 40 & 60 & 180 & 120 & 85 & 60\\
            \hline
        \end{tabular}
    }
    \label{table:DG}
\end{table}

\begin{table}[t!]
    \centering
    \caption{HT PIM Parameters}
    \scalebox{0.9}{
        \begin{tabular}{|c|c|c|c|c|c|c|}
            \hline
             \rowcolor{black!20} &  \parbox{0.75cm}{\centering $R^\mathrm{tf}$\\($\Omega$)} & \parbox{0.75cm}{\centering $X^\mathrm{tf}$\\($\Omega$)} & \parbox{0.75cm}{\centering $R^\mathrm{HT}$\\($\Omega$)} & \parbox{0.75cm}{\centering $X^\mathrm{HT}$\\($\Omega$)} & \parbox{0.75cm}{\centering $\overline{S}^\mathrm{tf}$\\(kVA)} &\parbox{0.75cm}{\centering $\overline{S}$\\(kVA)}\\
             \hline
            \cellcolor{black!10}HT 1 & 5.333$e^{-5}$ & 4.213$e^{-3}$ & 0 & 4.213$e^{-4}$ & 1000 & 100\\
            \hline
            \cellcolor{black!10}HT 2 & 3.2$e^{-3}$ & 1.28$e^{-2}$ & 0 & 1.32$e^{-3}$ & 166.67 & 16.67\\
            \hline
            \cellcolor{black!10}HT 3 & 3.2$e^{-3}$ & 1.28$e^{-2}$ & 0 & 1.32$e^{-3}$ & 166.67 & 16.67\\
            \hline
             \cellcolor{black!10}HT 4 & 5.3$e^{-3}$ & 2.13$e^{-2}$ & 0 & 2.19$e^{-3}$ & 100 & 10\\
            \hline
             \cellcolor{black!10}HT 5 & 1.78$e^{-3}$ & 7.11$e^{-3}$ & 0 & 7.33$e^{-4}$ & 300 & 30\\
            \hline
             \cellcolor{black!10}HT 6 & 3.56$e^{-3}$ & 1.422$e^{-2}$ & 0 & 1.47$e^{-3}$ & 150 & 15\\
            \hline
        \end{tabular}
    }
    \label{table:HT Parameters}
\end{table}

\begin{table}[t!]
    \centering
    \caption{HT and Transformer Costs}
    \begin{adjustbox}{width=1\linewidth}
        \begin{tabular}{|>{\columncolor{black!10}}c|c|c|c|c|c|c|}
            \hline
              \rowcolor{black!20} & HT 1 & HT 2 & HT 3 & HT 4 & HT 5 & HT 6\\
             \hline
            \cellcolor{black!10} HT Cost (£) & 78,260.00 & 24,8832.50& 24,8832.50& 20,317.50& 35,692.60& 23,297.40\\
            \hline
            \cellcolor{black!10} \Centerstack{Transformer \\Cost (£)} & 52,000.00 & 16,500.00 & 16,500.00 & 13,500.00 & 23,716.00 & 15,480.00 \\
            \hline
            \cellcolor{black!10}\Centerstack{Offset \\Cost (£)} & 26,000.00 & 8,332.50 & 8,332.50 & 6,817.50 & 11,976.60 & 7,817.40 \\
            \hline
        \end{tabular}
    \end{adjustbox}
    \label{table:HT costs}
\end{table}

For the Cigre benchmark portion of the test case network the default maximum loads presented in \cite{CIGRE2014} have been used, excluding the loads for buses R1 and C1. For Benchmarks A and B, the maximum loads were selected based on the recommendations in \cite{Dickert2010}, similar to what was done in \cite{Dickert2013}. For Benchmark A, buses A2-A13 have a 3-phase load of 10kVA, with a power factor of 0.95. For Benchmark B, all buses downstream of the LV transformer have a 3-phase load of 1.5kVA, with a power factor of 0.95.

\begin{figure}[t!]
    \centering
    \includegraphics[width=0.6\linewidth]{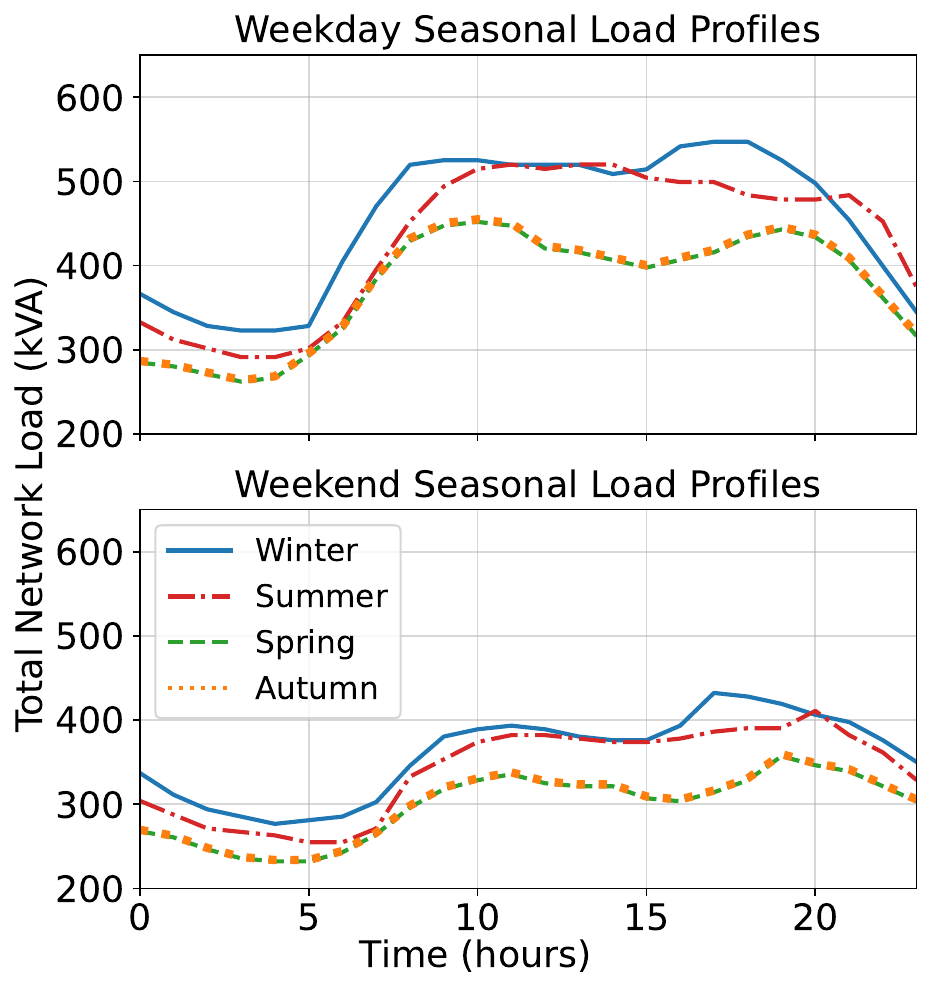}
    \caption{Seasonal load profiles used for the test case}
    \label{fig:Seasonal Loads}
\end{figure}

For this case study, load profiles for the average weekdays and weekends for all four seasons are used to approximate the NPV of HTs. For this case study, no specific year has been chosen, therefore, there is no exact number of weekdays or weekends. Instead, it is assumed that $5/7^{th}$ of each season is comprised of weekdays, and $2/7^{th}$ of weekends. The load profiles for the average weekday and weekend for each season are shown in Figure \ref{fig:Seasonal Loads}, where the total network load for each time period is shown. Each day is represented by 24 time periods, one per hour. For every time period, the maximum network load was multiplied by a modifier to obtain the network loads for that time period, similar to what was done in \cite{Hung2013}. The seasonal load profile modifiers were calculated from the values given in \cite{Grigg1999}.

HT LP and HT SLP parameters are presented in Tables \ref{table:OPF Parameters} and \ref{table:Weights}. As seen in Table \ref{table:Weights}, changes in energy export prices, HT costs, and discount rates would change the size and order of magnitude of the NPV calculated, as such different weight values were used for the different test case scenarios. $C^\mathrm{ref}$ values £97,310, £389,240.11, and £1,167,720.33 were used for energy export prices £0.01/kWh, £0.04/kWh and £0.12/kWh, respectively. All $C^\mathrm{ref}$ values were calculated from an alternative SLP method that had the objective of maximising the estimated annual profit made from exporting active DG power without utilising HTs. For these test cases, in the HT SLP `Network Feasibility Check', the network voltage per unit values were rounded to 2 decimal places. As such, no voltage violation 0.005 pu or greater was permitted. This decision was made as it would prevent significant voltage violations while disregarding very minor voltage violations, as previously mentioned in Section \ref{section:SLP}.

\begin{table}[t!]
    \renewcommand*{\arraystretch}{1.2}
    \centering
    \caption{Test Case HT LP and HT SLP Constants and Parameters}
    \begin{tabular}{|c|c|c|c|}
        \hline
        \cellcolor{black!10}$C_1$ & 0.99 & \cellcolor{black!10}$C_2$ & 0.001
        \\
        \hline
        \cellcolor{black!10}$v^\mathrm{pq\ tol}$ & 5\% & \cellcolor{black!10}$Obj^\mathrm{tol}$ & 1\%\\
        \hline
        \cellcolor{black!10}$\lvert \overline{v} \rvert$ & 1.1pu & \cellcolor{black!10}$\lvert \underline{v} \rvert$ & 0.9pu\\
        \hline
        \cellcolor{black!10}$v^\mathrm{base}$ & 230.94V & \cellcolor{black!10}$s^\mathrm{base}$ & 166.67 kVA \\
        \hline
        \cellcolor{black!10}$stp^{v}$ & 0.05 & \cellcolor{black!10}$stp^{\gamma}$ & 0.1$\pi$ \\
        \hline
        \cellcolor{black!10}$\underline{stp}_{i}^\mathrm{DG}$ & 0.005$\overline{P}_{i} \ \forall i \ \in \ \mathcal{N}^\mathrm{DG}$ & \cellcolor{black!10}$md^\mathrm{HT}$ & 100 \\
        \hline
        \cellcolor{black!10}$HT^\mathrm{yrs}$ & 20 & \cellcolor{black!10}$\overline{k}$ & 50 \\
        \hline
        \cellcolor{black!10}$stp_{i}^\mathrm{DG}$ & \multicolumn{3}{c|}{0.333$\overline{P}_{i} \ \forall i \ \in \ \mathcal{N}^\mathrm{DG}$} \\
        \hline
    \end{tabular}
    \label{table:OPF Parameters}
\end{table}

\begin{table}[t!]
    \centering
    \caption{Test Case HT LP Weight Values}
        \begin{tabular}{|c|c|c|c|c|}
            \hline
             \rowcolor{black!20} Test Cases & Weights & £0.01/kWh & £0.04/kWh & £0.12/kWh \\
             \hline
            \cellcolor{black!10}&$w_1$  & $91.25$ & $730$ & $9581.25$\\
            \cline{2-5}
            \cellcolor{black!10}&$w_2$  & $9.125$ & $73$ & $958.125$\\
            \cline{2-5}
            \cellcolor{black!10}\multirow{-3}{*}{Full-Cost}&$w_3$  & $91.25$ & $730$ & $9581.25$\\
            \Xhline{5\arrayrulewidth}
            \cellcolor{black!10}&$w_1$  & $228.125$ & $2737.5$ & $9581.25$\\
            \cline{2-5}
            \cellcolor{black!10}&$w_2$  & $22.8125$ & $273.75$ & $958.125$\\
            \cline{2-5}
            \cellcolor{black!10}\multirow{-3}{*}{Offset-Cost}&$w_3$  & $228.125$ & $2737.5$ & $9581.25$\\
            \hline
        \end{tabular}
    \label{table:Weights}
\end{table}

\subsection{Test Case Results}
A sensitivity analysis was completed for the case study to investigate the impact of different export prices, the discount rates on calculating the NPV, and the investment cost of the HTs. Export prices of £0.01/kWh, £0.04/kWh, and £0.12/kWh were used, and discount rates of 5\%, 7\%, and 10\%. In all test case scenarios, the energy export price was assumed to be constant throughout all time periods. The energy export prices used were based on the Smart Export Guarantee (SEG) \cite{Ofgem2024} prices offered by the energy suppliers E \cite{E_SEG2024}, OVO \cite{OVO2024}, and Scottish Power (for non-customers) \cite{ScotPower2024}. Two sets of HT investment costs were used for the test cases, one where the full costs of the HTs were used (referred to as the full-cost test cases), and another where HT costs were offset by the costs of conventional transformers with matching power and voltage ratings (referred to as the offset-cost test cases). The full-cost test cases represent the scenario in which the conventional network transformers that are being replaced cannot be sold to offset HT investment costs. The offset-cost test cases represent the scenario in which the conventional transformers that are being replaced are sold to offset the HT investment costs.

\begin{figure}[t!]
    \centering
    \includegraphics[width=0.6\linewidth]{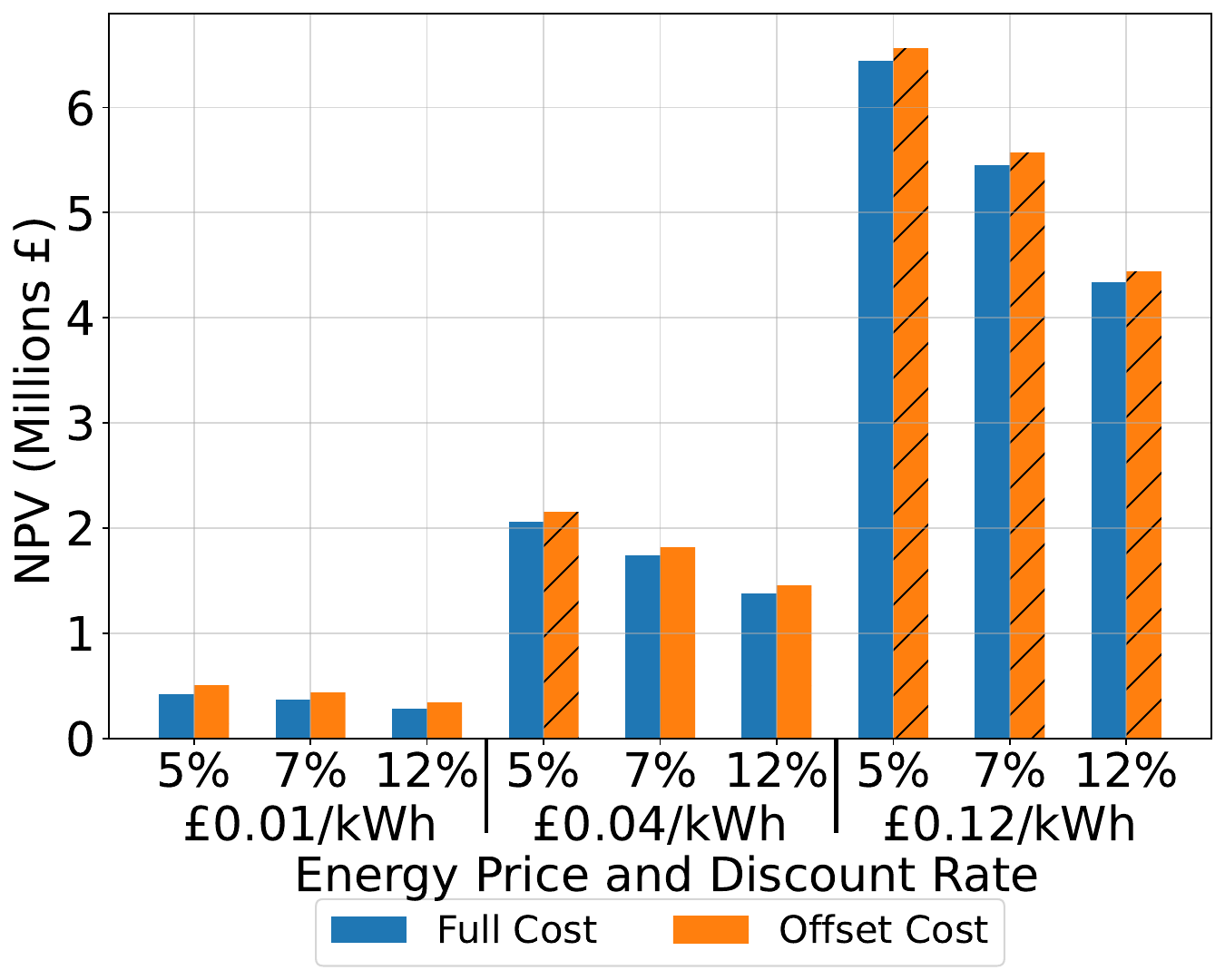}
    \caption{NPV Results for Full Cost and Offset Cost test cases}
    \label{fig:NPV_fig}
\end{figure}

\begin{figure}[t!]
    \centering
    \includegraphics[width=0.9\linewidth]{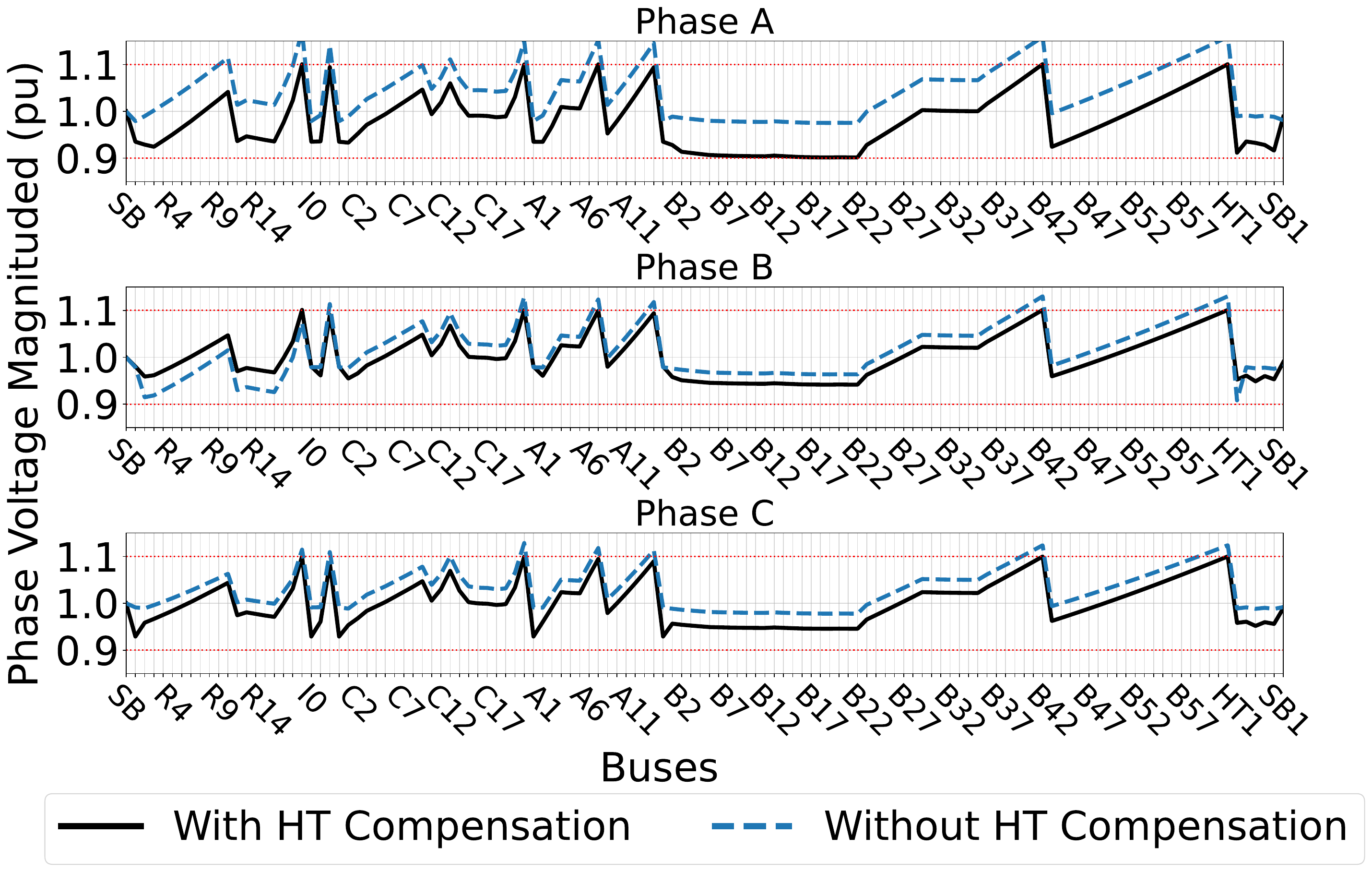}
    \caption{Comparison of network phase voltage magnitudes with and without HT compensation (full-cost test case, £0.12/kWh. 5\% discount rate, Spring weekend 4 a.m.)}
    \label{fig:Volt_Profiles}
\end{figure}

\begin{figure}[t!]
    \centering
    \includegraphics[width=1\linewidth]{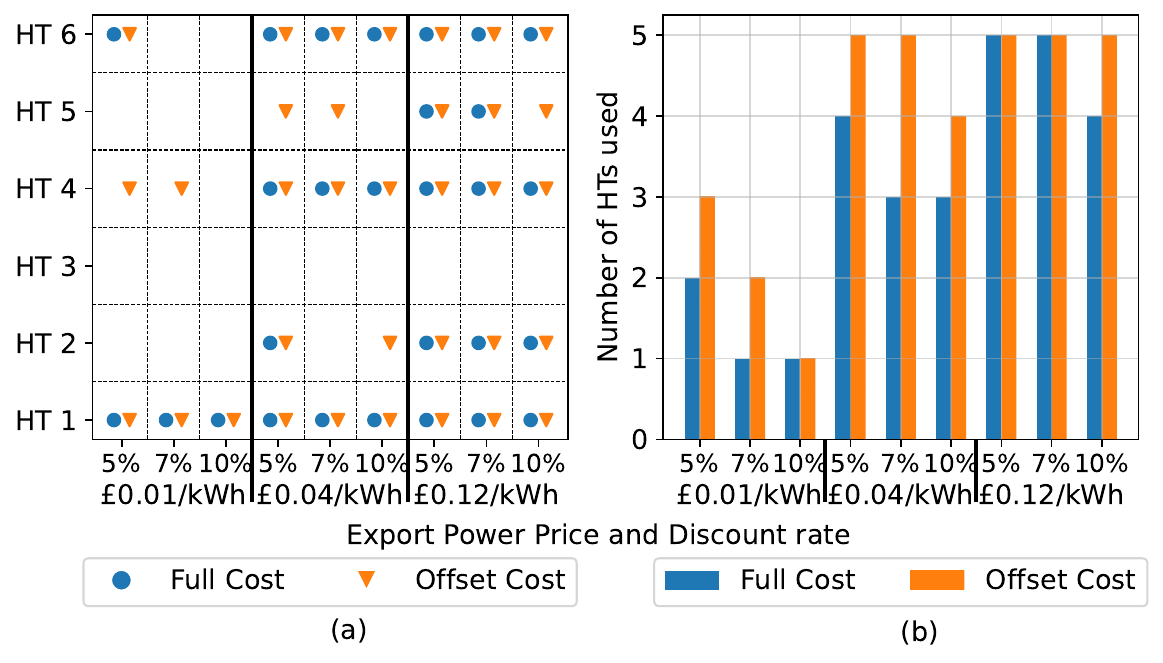}
    \caption{a). Combinations of HTs utilised, (b). Total number of HTs installed}
    \label{fig:HT_combo}
\end{figure}

The results for both the full-cost and offset-cost test cases can be seen in Table \ref{table:Results_NPV}, and Figures \ref{fig:NPV_fig}, \ref{fig:Volt_Profiles}, and \ref{fig:HT_combo}. In Table \ref{table:Results_NPV}, it can be seen that for all the full-cost and offset-cost test cases, HT installation results in a positive NPV, indicating increased power export justifies the investment costs. It can also be seen in Figure \ref{fig:HT_combo} that higher energy export prices, lower discount rates, and lower HT investment costs encourage more HTs to be installed. Compared to the full-cost test case results, the offset-cost test case results have greater NPVs and more HTs installed. The differences in NPV between the full-cost and offset-cost test cases are shown in Figure \ref{fig:NPV_fig} and Table \ref{table:Results_NPV}, with the greatest difference in percentage being a 19.81\% increase in NPV (for test cases with export price of £0.01 and 10\% discount rate).

In all cases, HT compensation was successfully utilised to prevent significant voltage violations. An example of this is shown in Figure \ref{fig:Volt_Profiles}, which compares network phase voltage levels with and without HT compensation for time period 28 (Spring Weekend 4 a.m.) for the full-cost test case with £0.12/kWh export price and 5\% discount rate.

In all cases, an optimal solution was found. The shortest HT SLP solve time was the offset-cost test case for the export price of £0.04/kWh and 7\% discount rate, which solved in 828 minutes and took 8 iterations. The longest HT SLP solve time was the full-cost test case for the export price of £0.04/kWh and 5\% discount rate, which solved in 1,375 minutes and took 12 iterations. Such solution times, while long, are reasonable for investment problems that span multiple years.

\begin{figure}[t!]
    \centering
    \includegraphics[width=0.65\linewidth]{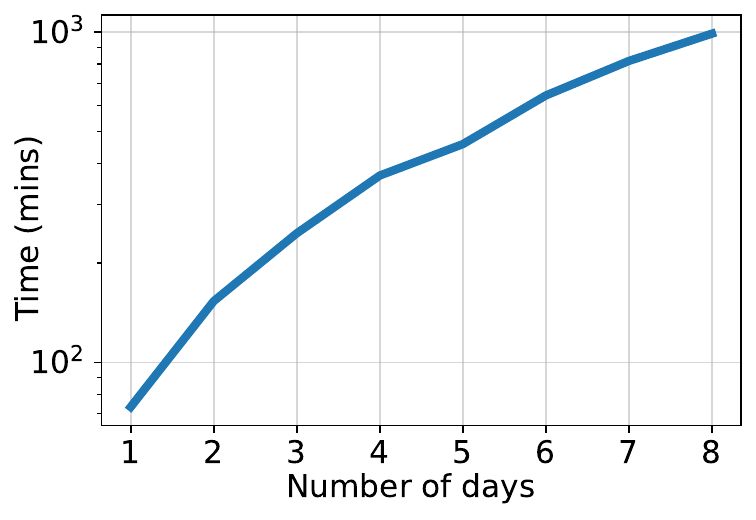}
    \caption{HT SLP solve time compared to the number of days}
    \label{fig:Time_v_day}
\end{figure}

In addition to the full-cost and offset-cost test cases, results were also gathered to determine how the number of time periods impacted the HT SLP solve time. A range of 1-8 days (24-196 time periods) was used for these results. When collecting these results, the same load profile (Summer weekday) was used for every day, so that the number of SLP iterations remained consistent between results. Otherwise, changes in the number of iterations would also impact the HT SLP solve time.It can be seen in Figure \ref{fig:Time_v_day} that as the number of time periods increases the HT SLP solution increases sub-exponentially.

\begin{table*}[t!]
    \centering
    \caption{Test Case NPV Results}
    \begin{tabular}{|c|c|c|c|c|c|}
        \hline
         \rowcolor{black!20} \makecell{Export Price \\ (£/kWh)} & \makecell{Discount Rate \\ (\%)} & \makecell{NPV (Full Cost)\\ (£)} & \makecell{NPV (Offset Cost)\\ (£)} & \makecell{NPV Difference\\ (£)} & \makecell{NPV Difference\\ (\%)} \\
         \hline
         \cellcolor{black!20}& \cellcolor{black!10}5 & 422,707.22 & 505,784.67 & 83,077.45 & 19.65 \\ \cline{2-6}
         \cellcolor{black!20}& \cellcolor{black!10}7 & 368,252.23 & 431,679.01 & 63426.78 & 17.22\\ \cline{2-6}
         \cellcolor{black!20}\multirow{-3}{*}{0.01}& \cellcolor{black!10}10 & 280,654.64 & 336,258.96 & 55,604.32 & 19.81\\
         \hline
         \cellcolor{black!20}& \cellcolor{black!10}5 & 2,059,627.36 & 2,148,488.50 & 88,861.14 & 4.31\\ \cline{2-6}
         \cellcolor{black!20}& \cellcolor{black!10}7 & 1,741,051.14 & 1,815,926.18 & 74,875.04 & 4.30 \\ \cline{2-6}
         \cellcolor{black!20}\multirow{-3}{*}{0.04}& \cellcolor{black!10}10 & 1,374,685.66 & 1,451,793.76 & 77,108.10 & 5.61\\
         \hline
         \cellcolor{black!20}& \cellcolor{black!10}5 & 6,443,875.12 & 6,564,187.22 & 120,312.10 & 1.87\\ \cline{2-6}
         \cellcolor{black!20}& \cellcolor{black!10}7 & 5,447,537.62 & 5,569,740.03 & 122,202.41 & 2.24\\ \cline{2-6}
         \cellcolor{black!20}\multirow{-3}{*}{0.12}& \cellcolor{black!10}10 & 4,339,445.81 & 4,440,684.14 & 101,238.33 & 2.33\\
         \hline
    \end{tabular}
    \label{table:Results_NPV}
\end{table*}

\section{CONCLUSION.}
\label{section:Conclusion}
This paper proposed an SLP method capable of placing, controlling, and coordinating HTs in unbalanced, 3-phase distribution networks. It has been shown that the HT SLP method is able to determine the optimal placement and coordination HTs which will increase the annual income earned from exporting excess DG active power output enough to justify HT investment costs, resulting in positive NPVs. Results also show that the nonlinearities of the HT PIMs are addressed by the proposed HT SLP method while maintaining an appropriate level of accuracy. Using the NPV results of this paper (the highest of which being £6.56 million, resulting from a 45.53\% increase in estimated annual profits due to coordinated HT compensation), the proposed HT SLP method has shown that the installation of HTs can be valuable investments. Potential future work that continues on from the HT SLP could include a multi-period SLP method that utilises HTs to help optimise the charging schedules of electric vehicles.


%
\ifCLASSOPTIONcaptionsoff
  \newpage
\fi



%
\bibliographystyle{IEEETran}
\bibliography{References}

\end{document}